\documentclass{aa}  

\usepackage{graphicx}
\usepackage{txfonts}
\usepackage{hyperref}
\usepackage{lscape}
\usepackage{longtable}
\hypersetup{breaklinks=true,colorlinks=true,linkcolor=blue,citecolor=blue,urlcolor=blue}

\graphicspath{{Figures/}}

\usepackage{xparse}

\usepackage[normalem]{ulem}

\NewDocumentCommand{\io}{s m m}{%
  \IfBooleanTF{#1}%
  {\MakeUppercase{#2}\,\textsc{#3}}%
  {[\MakeUppercase{#2}\,\textsc{#3}]}%
}
\newcommand{\ior}[3]{\io{#1}{#2}\,\textrm{$\lambda$}\,#3}

\newcommand{\ev}{4k$\times$4k E2V}

\newcommand{\uflux}{erg s$^{-1}$ cm$^{-2}$ \AA$^{-1}$}

\usepackage{listings}
\usepackage{xcolor}

\definecolor{codegreen}{rgb}{0,0.6,0}
\definecolor{codegray}{rgb}{0.5,0.5,0.5}
\definecolor{codepurple}{rgb}{0.58,0,0.82}
\definecolor{backcolour}{rgb}{0.95,0.95,0.92}

\lstdefinestyle{mystyle}{
    backgroundcolor=\color{backcolour},   
    commentstyle=\color{codegreen},
    keywordstyle=\color{blue},
    numberstyle=\tiny\color{codegray},
    stringstyle=\color{codepurple},
    basicstyle=\ttfamily\footnotesize,
    breakatwhitespace=false,         
    showspaces=false,                
    showstringspaces=false,
    showtabs=false,                  
    tabsize=2
}

\newcommand{\urltabs}{cavity.caha.es/data/dr1/}

\newcommand{\urls}[1]{\href{https://#1}{#1}}

\begin{document} 

   \title{CAVITY:\ Calar Alto Void Integral-field Treasury surveY}

   \subtitle{I. First public data release\thanks{Based on observations collected at the Centro Astron\'omico Hispano en Andaluc\'ia (CAHA) at Calar Alto, operated jointly by Junta de Andaluc\'ia and Consejo Superior de Investigaciones Cient\'ificas (IAA-CSIC).}}

   \author{Rubén García-Benito
          \inst{\ref{iaa}}
          \and Andoni Jiménez\inst{\ref{ugr1}}
          \and Laura Sánchez-Menguiano\inst{\ref{ugr1},\ref{ugr2}}
          \and Tomás Ruiz-Lara\inst{\ref{ugr1},\ref{ugr2}}
          \and Salvador Duarte Puertas\inst{\ref{ugr1},\ref{ugr2},\ref{laval}}
          \and Jesús Domínguez-Gómez\inst{\ref{ugr1}}
          \and Bahar Bidaran\inst{\ref{ugr1}}
          \and Gloria Torres-Ríos\inst{\ref{ugr1}}
          \and María Argudo-Fernández\inst{\ref{ugr1},\ref{ugr2}}
          \and Daniel Espada\inst{\ref{ugr1},\ref{ugr2}}
          \and Isabel Pérez\inst{\ref{ugr1},\ref{ugr2}}
          \and Simon Verley\inst{\ref{ugr1},\ref{ugr2}}
          \and Ana M.~Conrado\inst{\ref{iaa}}
          \and Estrella Florido\inst{\ref{ugr1},\ref{ugr2}}
          \and Mónica I. Rodríguez\inst{\ref{ugr1}}
          \and Almudena Zurita\inst{\ref{ugr1},\ref{ugr2}}
          \and Manuel Alcázar-Laynez\inst{\ref{ugr1}}
          \and Simon B. De Daniloff\inst{\ref{ugr1},\ref{iram}}
          \and Ute Lisenfeld\inst{\ref{ugr1},\ref{ugr2}}
          \and Rien van de Weygaert\inst{\ref{kapteyn}}
          \and Hélène M. Courtois\inst{\ref{cbl}}
          \and Jes\'us Falc\'on-Barroso\inst{\ref{iac},\ref{ull}}
          \and Anna Ferr\'e-Mateu\inst{\ref{iac},\ref{ull}}
          \and Llu\'is Galbany\inst{\ref{ice},\ref{ieec}} 
          \and Rosa M.~González Delgado\inst{\ref{iaa}}
          \and Ignacio del Moral-Castro\inst{\ref{chile}}
          \and Reynier F. Peletier\inst{\ref{kapteyn}}
          \and Javier Román\inst{\ref{ucm}}
          \and Sebastián F. Sánchez\inst{\ref{unam}}
          \and Pablo M.~S\'anchez-Alarc\'on\inst{\ref{iac},\ref{ull}}
          \and Patricia S\'anchez-Bl\'azquez\inst{\ref{ucm}}
          \and Pedro Villalba-González\inst{\ref{columbia}}
          %%% CAHA %%%
          \and Marco Azzaro\inst{\ref{caha}} 
          \and Martín Blazek\inst{\ref{caha}} 
          \and Alba Fernández\inst{\ref{caha}}
          \and Julia Gallego\inst{\ref{caha}}
          \and Samuel Góngora\inst{\ref{caha}} 
          \and Ana Guijarro\inst{\ref{caha}} 
          \and Enrique de Guindos\inst{\ref{caha}} 
          \and Israel Hermelo\inst{\ref{caha}} 
          \and Ricardo Hernández\inst{\ref{caha}}
          \and Enrique de Juan\inst{\ref{caha}} 
          \and José Ignacio Vico Linares\inst{\ref{caha}} 
          }

\institute{
Instituto de Astrof\'isica de Andaluc\'ia - CSIC, Glorieta de la Astronomía s/n, 18008 Granada, Spain\label{iaa}          
\and Dpto. de F\'{\i}sica Te\'orica y del Cosmos, University of Granada, Facultad de Ciencias (Edificio Mecenas), E-18071, Granada, Spain\label{ugr1}
%Dpto. de F\'\i sica Te\'orica y del Cosmos, Edificio Mecenas, Campus de Fuentenueva, Universidad de Granada, E18071, Granada, Spain
\and Instituto Carlos I de F\'\i sica Te\'orica y Computacional, Universidad de Granada, E18071, Granada, Spain\label{ugr2}
\and D\'epartement de Physique, de G\'enie Physique et d’Optique, Universit\'e Laval, and Centre de Recherche en Astrophysique du Qu\'ebec (CRAQ), Québec, QC, G1V 0A6, Canada\label{laval}
\and Institut de Radioastonomie Millim\'etrique (IRAM), Av. Divina Pastora 7, N\'ucleo Central 18012, Granada, Spain\label{iram}
\and Kapteyn Astronomical Institute, University of Groningen, PO Box 800, 9700 AV Groningen, The Netherlands\label{kapteyn}
\and Universit\'e Claude Bernard Lyon 1, IUF, IP2I Lyon, 4 rue Enrico Fermi, 69622 Villeurbanne, France\label{cbl}
\and Instituto de Astrof\'isica de Canarias, c/V\'ia L\'actea s/n, E-38205, La Laguna, Tenerife, Spain\label{iac}
\and Departamento de Astrof\'isica, Universidad de La Laguna, E-38206, La Laguna, Tenerife, Spain\label{ull}
\and Institute of Space Sciences (ICE-CSIC), Campus UAB, Carrer de Can Magrans, s/n, E-08193 Barcelona, Spain\label{ice}
\and Institut d'Estudis Espacials de Catalunya (IEEC), 08860 Castelldefels (Barcelona), Spain\label{ieec}
\and Instituto de Astrof\'isica, Facultad de F\'isica, Pontificia Universidad Cat\'olica de Chile, Campus San Joaqu\'in, Av. Vicu\~na Mackenna 4860, Macul, Santiago, Chile, 7820436\label{chile}
\and Departamento de Física de la Tierra y Astrofísica, Universidad Complutense de Madrid, E-28040 Madrid, Spain\label{ucm}
\and Instituto de Astronomía, Universidad Nacional Autonóma de México, A.P. 70-264, 04510 Ciudad de México, México\label{unam}
\and Department of Physics and Astronomy, University of British Columbia, Vancouver, BC V6T 1Z1, Canada\label{columbia}
\and Centro Astronómico Hispano en Andalucía, Observatorio de Calar Alto, Sierra de los Filabres, 04550 Gérgal, Almería, Spain\label{caha}
}

\date{Received 2024; accepted 2024}

% \abstract{}{}{}{}{} 
% 5 {} token are mandatory
 
  \abstract
  {The Calar Alto Void Integral-field Treasury surveY (CAVITY) is a legacy project aimed at characterising the population of galaxies inhabiting voids, which are  the most under-dense regions of the cosmic web, located in the Local Universe. This paper describes the first public data release (DR1) of CAVITY, comprising science-grade optical data cubes for the initial 100 out of a total of $\sim$300 galaxies in the Local Universe ($0.005~<~z~<~0.050$). These data were acquired using the integral-field spectrograph PMAS/PPak mounted on the 3.5m telescope at the Calar Alto observatory. The DR1 galaxy sample encompasses diverse characteristics in the color–magnitude space, morphological type, stellar mass, and gas ionisation conditions, providing a rich resource for addressing key questions in galaxy evolution through spatially resolved spectroscopy. The galaxies in this study were observed with the low-resolution V500 set-up, spanning the wavelength range 3745–7500 \AA,\ with a spectral resolution of 6.0 \AA\ (FWHM). Here, we describe the data reduction and characteristics and data structure of the CAVITY datasets essential for their scientific utilisation, highlighting such concerns  as vignetting effects, as well as the identification of bad pixels and management of spatially correlated noise. 
  We also provide instructions for accessing the CAVITY datasets and associated ancillary data through the project's dedicated database.}


   \maketitle
%
%-------------------------------------------------------------------

\section{Introduction}

The distribution of galaxies in the cosmos is heterogeneous, forming a complex structure known as the cosmic web. On megaparsec scales, this web is characterised by elongated filaments, sheet-like walls, and underdense voids \citep{Bond:1996,Weygaert:2008a,Weygaert:2008b}.
Despite occupying approximately 70\% of the Universe's present-time volume, voids contain only about 10\% of its mass \citep{Cautun:2014,Libeskind:2018}. These voids, originating from early density fluctuations, represent regions of weaker gravitational pull, expanding faster than the general Hubble flow and channeling matter into surrounding structures. Understanding void dynamics is pivotal for unraveling the overarching large-scale arrangement of the cosmic web and its impact on galaxy evolution \citep{Courtois:2023}.

Galaxies residing within voids provide unique insights into evolutionary processes, distinct from those in denser environments. They undergo different transformation mechanisms, offering clues into galaxy evolution dynamics. Void galaxies exhibit distinct characteristics, such as bluer colors and later morphological types \citep{Rojas:2004,Rojas:2005,Park:2007,Constantin:2008}, suggesting potential differences in their star formation rates. 

With the purpose of investigating systematic differences between void galaxies and those in denser environments, the Void Galaxy Survey  \citep[VGS;][]{Kreckel:2011, Kreckel:2012, Beygu:2016, Beygu:2017} has served as a multi-wavelength program focusing on 60 void galaxies in the Local Universe,  primarily utilising SDSS multiband photometry and HI data. Each galaxy was selected from the innermost regions of identified voids in the SDSS redshift survey using a geometric-topological watershed technique \citep{Platen:2007}, without any prior selection based on intrinsic properties of the void galaxies. The project comprehensively studied the gas content, star formation history, stellar content, as well as the kinematics and dynamics of void galaxies and their companions. One of the most intriguing findings of the VGS is the potential evidence for cold gas accretion in several noteworthy objects, including a polar ring galaxy and a filamentary arrangement of void galaxies \citep{Stanonik:2009}. 

However, the precise impact of void environments on various galaxy properties, including stellar mass assembly and metallicity content, especially on spatially resolved scales, remains contentious and requires further exploration, particularly through such methodologies as integral field spectroscopy (IFS).

The Calar Alto Void Integral-field Treasury SurveY (CAVITY) is a comprehensive project aimed at studying galaxies within cosmic voids in the Local Universe using IFS data \citep{Perez:2024}. Its primary focus lies in examining the spatially resolved stellar populations, stellar kinematics, and physical properties of ionised gas within these galaxies, alongside discerning their dark matter content, acknowledging the non-linear correlation between baryonic mass and dark halo. The CAVITY design ensures a thorough investigation of void galaxies that spans various void sizes and dynamical stages, while maintaining adequate spatial coverage across galaxy disks. 
Galaxy feedback and black hole growth mechanisms are highly dependent on the surrounding galaxy or halo environment, making them key challenges in cosmological numerical simulations. These processes, which influence star formation and galaxy evolution, vary significantly depending on factors like local density and interactions. The CAVITY project, by focusing on galaxy evolution in the lowest density regions of the Universe, will provide valuable insights into mass assembly and chemical enrichment in these environments. These results will help constrain the baryonic physics used in simulations, offering a better understanding of how feedback and black hole growth operate in low-density regions and improving $\Lambda$CDM model predictions.
Several scientific objectives have been explored using the CAVITY parent sample. In one study, \cite{Dominguez:2023a} utilised Sloan Digital Sky Survey (SDSS) spectra to analyze a well-defined group of galaxies in voids, walls and filaments, and clusters. By applying non-parametric full spectral fitting techniques, they measured the stellar metallicity in the central regions of these galaxies as covered by the SDSS fiber. They found that void galaxies possess lower stellar metallicities compared to those in filaments and walls, particularly among galaxies with lower stellar mass. Another study by \cite{Dominguez:2023b}, using the same methodology, examined the star formation histories (SFH) of galaxies in various environments. Their findings indicated that, on average, galaxies in voids assemble mass more slowly at a given stellar mass than galaxies in filaments or clusters.

In a more recent investigation utilising IFS spatially resolved data from the CAVITY survey, \cite{Conrado:2024} found that void galaxies generally have a slightly larger half-light radius (HLR), lower stellar mass surface density, and younger ages across all morphological types compared to a sample of galaxies in filaments and walls observed with the same instrument selected from the CALIFA survey. Additionally, void galaxies exhibited slightly higher star formation rates (SFR) and specific star formation rates (sSFR), especially in Sa-type galaxies. These differences were most significant in the outer regions of spiral galaxies, where the discs appeared younger and had higher sSFR, suggesting less evolved discs. Early-type spirals in voids also demonstrated a slower transition from star-forming to quiescent states. Overall, the study concluded that void galaxies evolve more gradually, particularly in their outer regions, with environmental effects having a more pronounced impact on low-mass galaxies.

In this article, we introduce the first data release (DR1) of CAVITY, which provides public access to 100 high-quality data cubes. Initially conceived as an IFS survey, CAVITY has undergone expansion to amplify its scientific impact and comprehensively address its proposed objectives. This expansion, named CAVITY+, entails the inclusion of additional observational techniques such as HI, CO, and deep optical imaging of the IFS targets. The current data release focuses solely on IFS data. 

This paper is organised as follows. The properties of the galaxies in the DR1 sample are summarised in Sect. \ref{sec:sample}. We describe the observing strategy and set-up (Sect. \ref{sec:obs}), processing (Sect. \ref{sec:reduc}), data format (Sect. \ref{sec:format}), and data quality (Sect. \ref{sec:qc}), which comprise essential information for any scientific analysis of the distributed CAVITY data. Access to the CAVITY DR1 data is explained in Sect. \ref{sec:access}.

\section{The CAVITY DR1 sample\label{sec:sample}} 

The CAVITY project focuses on identifying and characterising cosmic voids in the nearby Universe, drawing its sample from the SDSS data cataloged by \cite{Pan:2012}. Ensuring completeness within the redshift range of 0.005 to 0.050 allows for the inclusion of voids spanning large sizes. From this initial void catalog \citep{Pan:2012}, and taking into consideration the scientific goals of the CAVITY project \citep{Perez:2024}, we established a criterion of at least 20 galaxies per void, culminating in 80 voids containing 19,732 galaxies. Subsequent refinement excluded voids at the edges of the SDSS footprint, leaving 8690 galaxies within 42 voids. Representative voids were then chosen to cover a spectrum of properties, ensuring observability from the Calar Alto Observatory while maintaining the distribution of key galaxy properties. This resulted in a final sample of 4866 galaxies in 15 voids, termed the CAVITY parent sample. Additionally, to guarantee sample purity, galaxies associated with clusters were meticulously removed. A criterion based on the distance to the center of the voids in units of the void effective radius (within the inner 0.8 $\times$ void effective radius) was employed to prioritise galaxies in the innermost regions of voids, covering the full extent of them while avoiding filamentary-like environments. Also, only galaxies with intermediate inclinations (20 to 70 degrees) were selected for observations. This led to a final observable sample of 1115 galaxies after an extra visual refinement process \citep{Perez:2024}. Target galaxies are constantly selected based on observability of galaxies from this parent sample. Since the beginning of the observations, the project has made use of 221 nights distributed between January 2021 and June 2024, collecting 246 cubes. This initial data release comprises 100 galaxies, chosen based on rigorous quality control measurements (Sect.~\ref{sec:qc}).

\begin{figure*}
    \includegraphics[width=0.95\columnwidth]{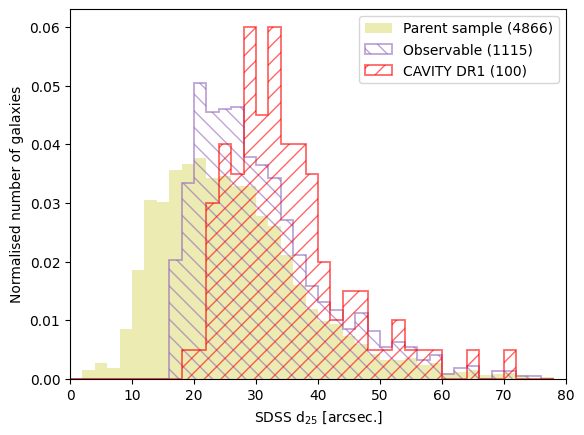}
    \includegraphics[width=0.93\columnwidth]{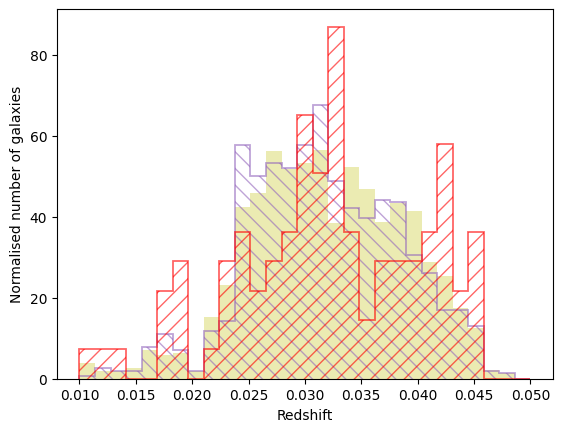}\\
    \includegraphics[width=0.95\columnwidth]{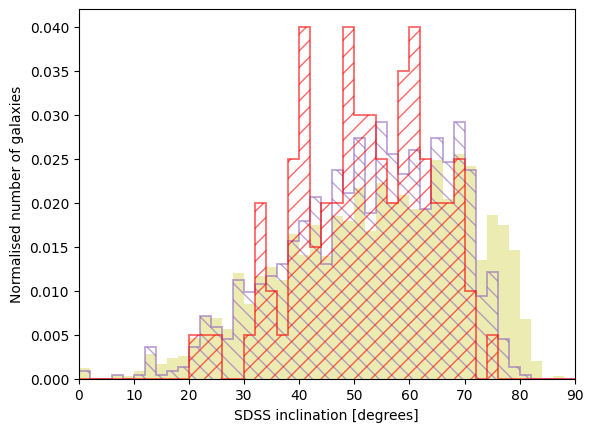}
    \includegraphics[width=0.93\columnwidth]{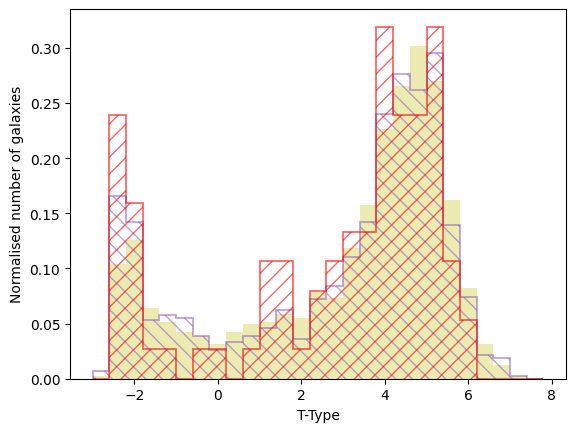}\\
    \includegraphics[width=0.95\columnwidth]{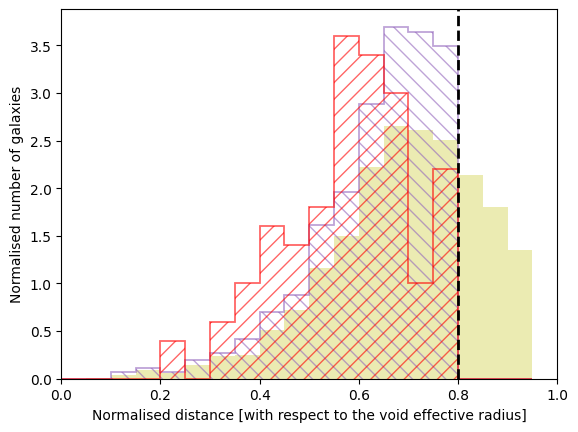}
    \includegraphics[width=0.93\columnwidth]{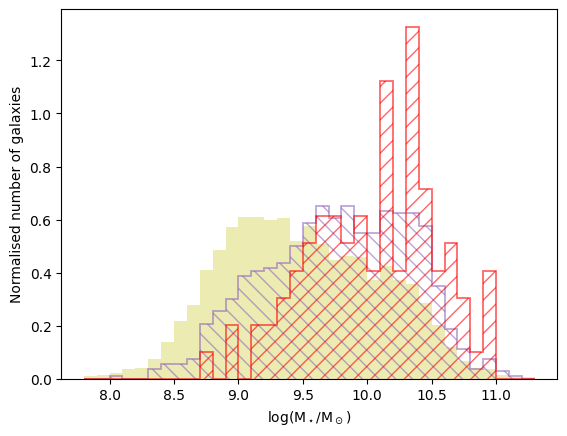}

\caption{Characterisation of the DR1 CAVITY sample in comparison with the CAVITY parent sample and observable sample that observed galaxies are drawn from (see text for details). From top to bottom and left to right, we show the distribution of d$_{25}$ (see text for definition), the stellar mass, the inclination from SDSS, the morphological T-Type, the normalised distance to the center of the voids, and the redshift. In light-green are the histograms corresponding to the CAVITY parent sample, in violet the observable sample, and in red the final CAVITY DR1 sample. The right part with respect to the black vertical dashed line in the normalised distance to the centre of the void panel histogram (bottom-left) represents the most external parts of the void (area between 0.8 and 1.0 times the void effective radius). This region is avoided as it can be considered as a transition zone from the void to the filamentary-like environment.}  \label{fig:sample}
\end{figure*}

Figure~\ref{fig:sample} provides a brief overview of the main characteristics of the three different CAVITY samples (i.e., parent sample in light-green, observable sample in violet, and the final DR1 sample in red). In order to ensure a 2D analysis of the properties of the galaxies, only those with  $d_{25}$ above $\sim$20\arcsec\ were observed (see top panel), with all galaxies being fully covered by the PMAS/PPak FoV (around 70 arcsecs). Also, $d_{25}$ has been computed as three times the value of its {\it petroR90-r}  ({\it r}-band) from SDSS\footnote{See `master' table description at the project webpage: \urls{\urltabs}}. Also, to allow for a proper study of the stellar and gas kinematics of the galaxies, systems with edge-on and face-on orientations are avoided. Thus, CAVITY DR1 galaxies exhibit only intermediate inclinations, with values ranging from $\sim$ 20 to 70 degrees (middle panel of Fig.~\ref{fig:sample}, by definition), with the bulk of galaxies displaying inclinations between 50 and 60 degrees. Finally, besides observability, galaxies to be observed as part of CAVITY are sorted according to the distance to the center of the void, prioritising observations of galaxies near the centers first. The number of galaxies within voids greatly decreases as we approach their cores, however, to maximise the scientific impact of the project, we aimed at covering all "void-centric" distances up to 0.8 void effective radius\footnote{Computed as the radius of a sphere with the same volume as the actual void.}, where the void environment gives way to a more filamentary-like surrounding. The distribution of distances to the void centers is shown in the bottom panel of Fig.~\ref{fig:sample}. For completeness, the figure also includes the distribution of redshifts, along with the distributions of morphological types \citep[extracted from][]{2018MNRAS.476.3661D} and stellar mass. We can see that the three samples cover a similar range of values in the redshift and morphological-type space, whereas a clear deficiency of low-mass galaxies is appreciated for the DR1 sample. This is mainly due to the technical difficulty of gathering high-quality IFS data for these faint low-mass (and generally small) galaxies. All these quantities are collected and explained in detail in the CAVITY DR1 "master" table (see Table~\ref{tab:CAVITY_master_table}). 

\begin{figure}    
    \includegraphics[width=\columnwidth]{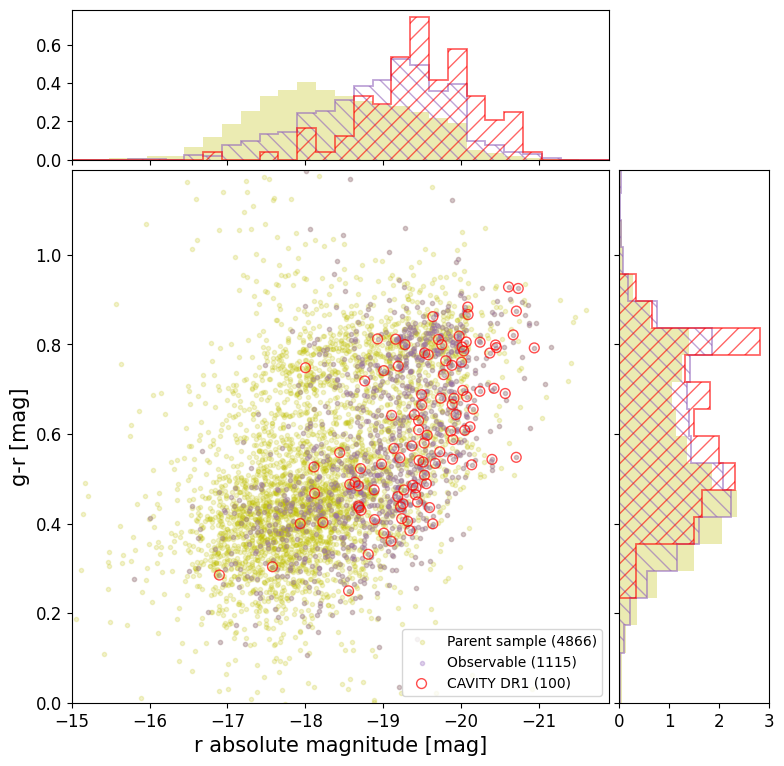}
\caption{Distribution of CAVITY galaxies in the SDSS g$-$r vs. M$_r$ color-magnitude diagram. Yellow dots (histograms) denote galaxies in the CAVITY parent sample (4886 galaxies), while violet and red are used for the observable subsample (1115) and the final CAVITY DR1 sample (100). Color and absolute magnitude distributions are displayed in the top and side panels.}  \label{fig:sample_CMD}
\end{figure}

Figure~\ref{fig:sample_CMD} shows the distribution of galaxies in the color-magnitude diagram (using SDSS magnitudes). The CAVITY DR1 sample mostly follows the overall distribution of the parent and observable samples (see also morphological type panel in Fig.~\ref{fig:sample}), with the exception of the very faint tail (galaxies fainter than absolute magnitude in {\it r}-band of -18 hardly enters the DR1 sample). Apart from this, we should highlight a subtle excess of galaxies in the bright end of the population (brighter than M$_{r}$=-20). This current lack of faint galaxies and excess of bright galaxies is translated into a slightly unbalanced stellar mass distribution (last panel of Fig.~\ref{fig:sample}). We will consider this in the planning of the coming observations, prioritising the observations of fainter systems.  

\begin{figure*}
 \includegraphics[width=\textwidth]{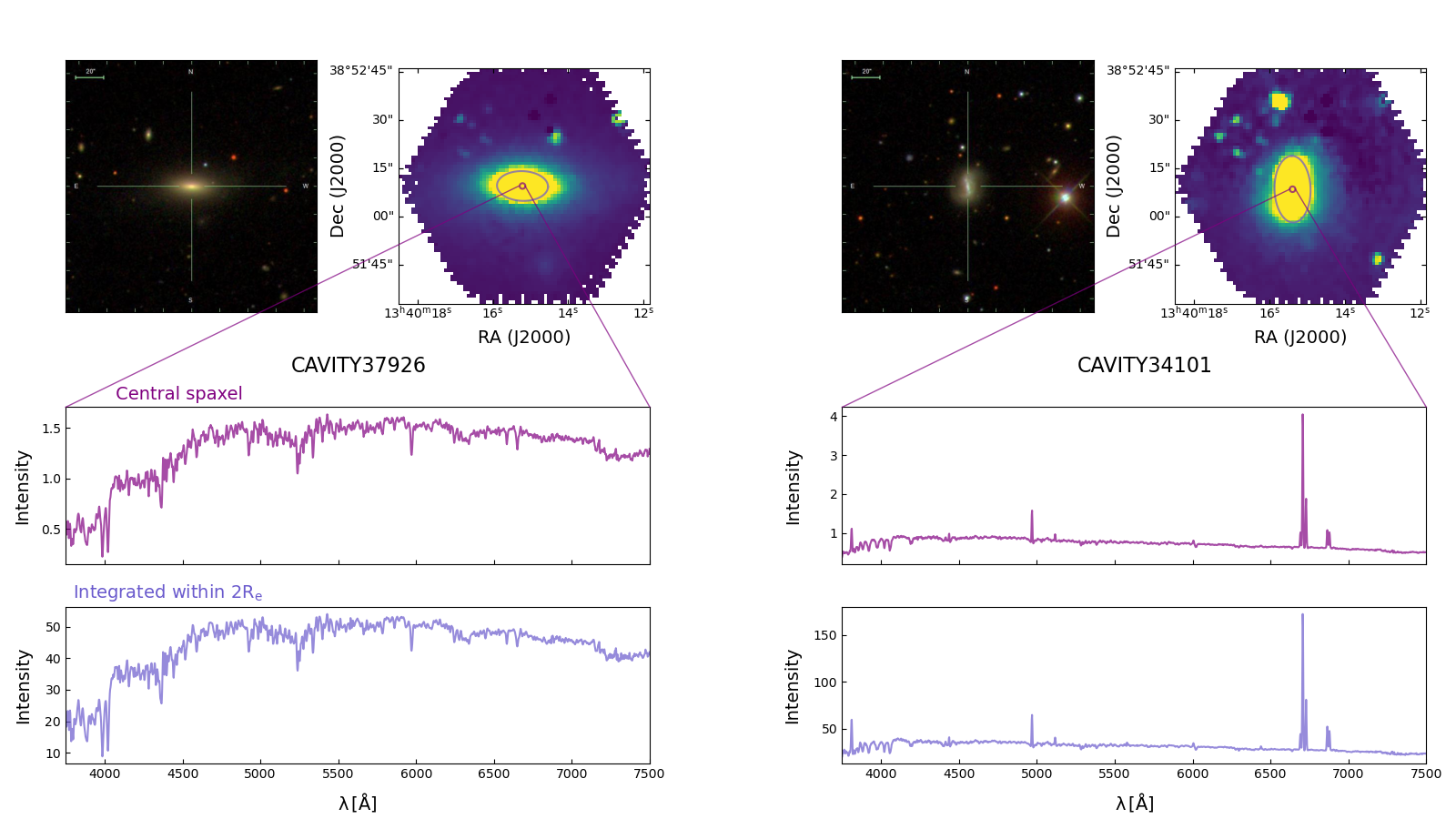} \\
\caption{Spectra of two representative CAVITY DR1 galaxies: CAVITY37926 (a red, quiescent galaxy; left column), and CAVITY34101 (a blue, star-forming galaxy; right column). \emph{Top-left:} SDSS color image of the galaxies. \emph{Top-right:} Moment 0 image reconstructed from the CAVITY data cube summing all the integrated light within the covered wavelength range (from 4500 to 7000 \AA). \emph{Middle} and \emph{bottom:} Spectra of their central spaxels and the integrated spectra up to 2 effective radii (violet ellipse in the moment 0 images), respectively. Moment 0 images are in units of 10$^{-16}$·erg·cm$^{-2}$·s$^{-1}$, and spectra in units of 10$^{-16}$·erg·cm$^{-2}$·s$^{-1}$·\AA$^{-1}$.
}
  \label{fig:spectra}
\end{figure*}

Figure~\ref{fig:spectra} presents some spectra belonging to two representative CAVITY DR1 galaxies. The left-hand panels correspond to a red, quiescent galaxy, and the right-hand ones to a blue, star-forming galaxy. For each galaxy we show in the top panels its SDSS color image (left) as well as its integrated light within the covered wavelength range extracted from the CAVITY data cube (right). We represent the spectra of their central spaxel in the middle panels and the integrated spectra up to 2 effective radii\footnote{The effective radius is obtained as the Petrosian half-light radius R50 measured in $r$-band and extracted from the MPA-JHU Catalogue \citep{2004Brinchmann}.} (violet ellipse in the images) in the bottom panels. This figure highlights the richness in physical properties covered by the CAVITY sample, as well as exemplifies the quality of the CAVITY DR1 dataset.

\section{Observing strategy and set-up overview\label{sec:obs}}

The observations of the CAVITY survey commenced in January 2021, conducted at the 3.5 m telescope of the Calar Alto observatory using the Potsdam Multi Aperture Spectrograph, PMAS \citep{Roth:2005}, in the PPak mode \citep{Verheijen:2004,Kelz:2006}. The PPak fiber bundle comprises 382 fibers, each with a diameter of 2.7\arcsec. Among these fibers, 331 (referred to as the science fibers) are concentrated in a single hexagonal bundle covering a field-of-view (FoV) of 74\arcsec $\times$ 64\arcsec, achieving a filling factor of approximately 60\%. The sky background is sampled by an additional set of 36 fibers, distributed in six bundles of six fibers each, positioned along a circle approximately 72\arcsec\ from the center of the instrument's FoV. These sky fibers are interspersed among the science fibers within the pseudo-slit to ensure a comprehensive characterisation of the sky, sampled with a similar distortion as the science fibers, while the remaining 15 fibers are utilised for calibration purposes. 

All DR1 CAVITY galaxies are observed using the V500 grating. It offers a nominal resolution of $\lambda/\Delta\lambda$ $\sim$ 850 at $\sim$ 5000 \AA\ with a full width at half maximum (FWHM) of about 6 \AA. Paired with the PMAS CCD E2V CCD231 4k$\times$4k, installed in 2009 \citep{Roth:2010}, it spans nominally from \ior{o}{ii}{3727} to \ior{s}{ii}{6731} in the rest frame for all objects in the survey. This wavelength coverage facilitates the investigation of the primary scientific objectives of the project, through the examination of stellar populations and the characteristics of the ionised gas, thanks to the diverse range of spectral features and emission line species available in the spectra.

The optical system exhibits some vignetting at the four corners of the CCD. This vignetting effect is illustrated in fig. 4 of \cite{Sanchez:2012}. It is characterised by a decrease in transmission at the four edges of the map. Approximately 30\% of the spectra near the edge of the detector experience some efficiency loss, which gradually worsens towards the corners of the CCD. Vignetting impacts no more than a quarter of the full spectral range, severely affecting less than 15\% of the fibers and less than 25\% of their wavelength range. The useful wavelength range varies in those fibers affected by vignetting, with the worst cases being reduced to 4240 to 7140 \AA. These effects manifest in the FoV as an annular ring at approximately $\sim$15\arcsec\ from the center of the FoV (see Fig. 4, right panel of \citealp{Sanchez:2012} and Fig. 11 of \citealp{Husemann:2013}). 

To ensure complete coverage of the central bundle's FoV and enhance spatial resolution, a dithering scheme employing three pointings has been implemented. This approach, akin to the one employed in the CALIFA survey \citep{Sanchez:2016}, conducted with the same instrumental set-up, is aimed at mitigating the vignetting effect. The scheme utilises wider offsets in RA and Dec: 0.00\arcsec, 0.00\arcsec; -5.22\arcsec, -4.84\arcsec; and -5.22\arcsec, +4.84\arcsec. By adopting this strategy, any spectrum affected by vignetting in the final dithered and rebinned dataset benefits from compensation. This is achieved by incorporating spectra from at least two adjacent, unvignetted fibers positioned closer than 2\arcsec.

The exposure times were determined based on our previous pilot project of void galaxies using the instrument. These times remain fixed for all targeted objects, which have been categorised into two groups: bright and faint galaxies. To categorise galaxies as faint or bright systems, we study elliptical isophotal profiles from $r$-band, DR12 SDSS photometric data. Galaxies with a surface brightness fainter (brighter) than 22.1 mag/arcsec$^2$ at \verb|petroR50_r| are considered faint (bright) galaxies.
Exposure times for the V500 range between 1.5 and 3.0 hours of total integration, depending on the galaxy brightness. Each dithering position involves two exposures of 900 seconds for the brightest targets, while four 900-second exposures are taken for the faintest galaxies. For the 1.5-hour observations, one set of calibration files is obtained per galaxy. For longer total integration times, a minimum of two sets of calibration exposures are acquired. All observations are conducted at airmasses below 1.4.

Given the simultaneous operation of multiple projects at the observatory on the same telescope, the allotted time for each project within a single night may be fractional. Consequently, due to the relatively lengthy exposure times per galaxy, certain galaxies may be observed across multiple nights. To minimise uncertainties related to positioning and transparency changes between nights, we enforce the completion of a single dithering position within the same night. For faint galaxies, this entails gathering at least the four 900-second exposures of one pointing and their corresponding calibration files on the same night. As a result, a galaxy may be observed within a single night, two nights, or three nights. Approximately 51\% of the galaxies are observed entirely within one night, 42\% within two nights, and the remainder over three nights, with one pointing conducted each night. Pointings that do not meet minimum quality control standards (see Sect. \ref{sec:qc}) are rescheduled and re-observed to ensure complete observation. These pointings are re-observed entirely, with either two or four single exposures. In the following section, we detail the primary stages of the CAVITY data reduction pipeline utilised in generating the DR1 data.

\section{Data processing\label{sec:reduc}}

The CAVITY data reduction process relies on an automated pipeline that operates independently, requiring minimal human intervention. It produces scientifically significant frames along with quality control metrics to assess the precision of the processed data. The CAVITY pipeline (V1.2) adheres to the fundamental reduction procedures and protocols outlined in the CALIFA survey \citep{Sanchez:2012,Husemann:2013}. Tailored to accommodate the specific requirements of the CAVITY survey, it incorporates novel architecture and efficiency enhancements.  Implemented in Python 3, the pipeline leverages Cython code \citep{Behnel:2011} to address select computationally intensive tasks. The following outlines the current implementation of the data reduction process.

\subsection{CCD readout procedure and frame combination\label{sec:ccd}}

The reduction workflow encompasses the propagation of Poisson and read-out noise, cosmic rays, defective CCD columns, and vignetting effects. As noted, the CCD employed since October 2009 comprises a \ev\ detector. The read-out software for the CCD stores each frame in four distinct FITS files, corresponding to the four amplifiers of the detector, each characterised by varying bias and gain levels. The standard procedure since then has been focussed on the four-amplifier read mode, generating four individual files, each with its own bias level and gain. However, in December 2021, CCD quadrant `a' exhibited abnormally higher readout values, with approximately 2.5 times more noise and a reduction of 2.5 times in counts in well-illuminated fibers. 
Consequently, the CCD dewar had to be unmounted and shipped to  Leibniz-Institut f\"{u}r Astrophysik Potsdam (AIP) for intervention. Suboptimal cable routing, leading to stress, was rectified, and loose connections were addressed during the intervention. Additionally, the read-out procedure was modified to the 2-amplifiers `bc' mode, which demonstrated better stability and performance post-fixing. The CCD dewar was subsequently returned and reinstalled in the instrument by the end of June 2022. During the commissioning and testing phase, various observations were conducted to ensure optimal performance. However, these observations conducted during commissioning did not meet the quality standards required for the survey. Consequently, the official resumption of CAVITY observations occurred in November 2022, following confirmation of system stability and optimal conditions. Thus, observations conducted before December 2021, prior to the CCD failure, were processed using the ``classic'' 4-amp read mode, which had been in use for over a decade. Meanwhile, observations were carried out after November 2022 were carried out using the 2-amp `bc' mode, which has since become the standard mode for CAVITY.

Initially, the four (prior to 2022) or two (after 2022) distinct FITS files from the \ev\ detector are merged into individual frames for each exposure. Subsequently, each frame undergoes pixel cleansing, which includes cosmic ray removal facilitated by PyCosmic \citep{Husemann:2012}.

\subsection{Spectral extraction\label{sec:spext}}

The PMAS instrument is susceptible to flexures \citep{Roth:2005}, which induce slight shifts of the signal on the CCD along the dispersion and cross-dispersion directions depending on the hour angle of the telescope. To mitigate this effect, calibration frames are acquired within a maximum range of 1.5 hours from the science frames. For longer exposures of low-luminosity galaxies, two calibration files are obtained. These calibration files are paired with science exposures for individual frames based on their temporal proximity. This approach ensures that, in most instances, the flexure pattern of the calibration and science frames is mitigated. Relative flexure offsets are assessed concerning the continuum and arc-lamp calibration frames, and the wavelength solution is adjusted for each individual science frame. The wavelength solution of sky lines is assessed against expected values, with any discrepancies attributed to flexure effects and corrected for individual science frames. These offsets exhibit small values, typically $\leq$ 0.5 pixels. 

The straylight map is reconstructed from fiber trace gaps and subtracted from calibration and science exposures. Fiber profile widths are determined using continuum lamp fiber positions for spectral extraction, with some fibers exhibiting over 15\% stray-light contribution depending on throughput and input signal, impacting object signal-to-noise (S/N) ratio. Reconstruction involves smoothing data with a 20-pixel median filter and fitting sixth-order polynomials to cross-dispersion slices, followed by 2D Gaussian kernel smoothing to suppress high-frequency structure. Straylight subtraction precedes fiber spectrum extraction, ensuring proper inclusion of stray-light effects in the data error budget.

Although adjacent spectra on the CCD can map to different locations in the spatial plane \citep{Kelz:2006}, the cross-talk, when employing a modified version of the Gaussian suppression \citep{Sanchez:2006}, is minimised to less than 1\%. Fiber profile widths are determined by averaging over blocks of 20 fibers, employing $\chi^2$ minimisation with integrated fiber fluxes and a common FWHM per block as free parameters, with fixed fiber positions derived from continuum lamp data, including flexure offset adjustments. These measured widths are then interpolated along the entire dispersion axis using a 5th-order polynomial. Spectrum extraction utilises the optimal extraction algorithm \citep{Horne:1986}, with Gaussian fiber profile positions and widths fixed based on previous measurements. Optimal extraction allows for error propagation based on individual pixel errors and exclusion of bad pixels. However, the flux extraction may be unreliable if all three central pixels of a fiber are flagged as bad. This would warrant a flagging of the corresponding spectral resolution element to prevent potential data artifacts.

The flux extracted for each pixel along the dispersion direction is stored in a row-stacked-spectrum file (RSS). Wavelength solution and spectral resolution variations are determined for each fiber using HeHgCd calibration lamp exposures within each calibration dataset. This data is employed to resample spectra onto a linear wavelength grid, incorporating flexure offsets in the wavelength solution explicitly to obviate an additional resampling step. Spectral resolution is homogenised using an adaptive Gaussian convolution. Error propagation during Gaussian convolution is performed analytically, while a Monte Carlo method is employed to estimate noise vector post-spline resampling. To address bad pixels, a two-pixel expansion of the bad pixel mask along the dispersion axis is implemented, followed by setting error values of bad pixels to a high value ($\sim 10^{10}$ counts) and replacing data with linear interpolation from the nearest unmasked pixels. Relative fiber-to-fiber transmission differences are addressed by comparing wavelength-calibrated science frames with twilight sky exposures. Additionally, pixels exhibiting a transmission drop below 70\%, are masked.

\subsection{Flux calibration}

The flux calibration was conducted using a master sensitivity curve derived from dozens of observations performed prior to the CAVITY observations, following the one adopted in \cite[Sect. 4.2]{Garcia-Benito:2015}. For future CAVITY DRs, we plan to compute sensitivity curves corresponding to the years of observations within the project to generate an updated sensitivity function. This approach will enable us to assess any differences compared to the previous computed curve, particularly in light of changes in the readout mode of the CCD (as described in Sect. \ref{sec:ccd}) and any evolution in fiber transmission.

To facilitate this future analysis, each night, we observed spectrophotometric standard stars listed in the Oke Catalogue\footnote{BD+25d4655, BD+28d4211, BD+33d2642, Feige 34, and HD 93521, are among the most commonly observed ones. Their corresponding flux-calibrated spectra are publicly accessible on the CAHA's webpage: \url{http://www.caha.es/pedraz/SSS/sss.html}.} \citep{Oke:1990}. Whenever feasible, we observed two different calibration stars on the same night. To mitigate potential flux losses stemming from the filling factor of PPak, instead of utilising a single pointing, we observed stars using the same dithering pattern as for galaxies.
This strategy aims to optimise the sensitivity across the wavelength range while addressing flux losses inherent to PPak's filling factor when using one single pointing \citep{Garcia-Benito:2009,Garcia-Benito:2010}.

The science RSS files for individual frames from each pointing are corrected for atmospheric extinction across the wavelength range. This correction takes into account the airmass and the monitored V band extinction, utilising the extinction measured by the Calar Alto Visual Extinction monitor (CAVEX) at the moment of the observations and the average extinction curve for the observatory \citep{Sanchez:2007}. Subsequently, the sensitivity curve is applied to these corrected frames, converting them from observed counts to intensity, thereby ensuring flux calibration (see Sect. \ref{sec:specphot} for details on the accuracy of the spectrophotometric calibration).

\subsection{Sky subtraction\label{sec:sky}}

Following flux calibration, frames undergo correction for telluric lines and sky subtraction. PPak is equipped with 36 fibers for sampling the sky, arranged around the science fiber-bundle in six small bundles of six fibers each, positioned approximately 75\arcsec\ from the center of the FoV \citep{Kelz:2006}. Objects selected for the CAVITY sample occupy only a portion of the FoV of the central PPak bundle, ensuring that all sky fibers remain unaffected by emission from the corresponding target. A sky spectrum is generated by combining the 36 sky fibers of PPak. To mitigate issues arising when a bright field star or neighboring galaxy fills an entire sky fiber bundle, we calculate the median of the 30 faintest sky fibers. Each derived night-sky spectrum is then used to determine the night-sky brightness by convolving the spectrum with the transmission curves of the Johnson V-band filter. This value serves to monitor the actual conditions during data acquisition.

\subsection{Cube reconstruction\label{sec:cube}}

The individual frames for each pointing are consolidated into a single pointing RSS file. To address potential variations in atmospheric transmission during the science exposures, these three pointings are standardised to a common intensity and response function. This standardisation is achieved by comparing the integrated spectra of the fibers within a 30\arcsec\ diameter aperture. Subsequently, the science spectra corresponding to the three dithered exposures are merged into a single frame comprising 993 spectra. The individual position tables corresponding to the PPak central bundle are combined using the offsets provided for the dither. Finally, the spectra undergo correction for Galactic extinction \citep{Schlegel:1998,Cardelli:1989}.

In this context, in the current implementation of the CAVITY pipeline for the DR1, we depart from other registration methods such as the one detailed in \cite{Garcia-Benito:2015}, which was first implemented in CALIFA DR2. In that registration approach, the flux for the 331 fiber apertures in each pointing is derived from sky-subtracted SDSS broad-band images covering the wavelengths of the observation. These apertures undergo shifts in both right ascension and declination within a search box centered on the nominal pointing coordinates. The optimal registration is determined by comparing $\chi^2$ values between the broad-band aperture-matched fluxes and those obtained from the RSS spectra. The photometric scale factor identified at the best-matching position is then utilised to recalibrate the absolute photometry of each individual RSS pointing. 

This approach entails that all IFS observations are accompanied by equivalent broad-band counterpart images for all galaxies. This is not problematic for CAVITY, as its sample is drawn from SDSS, where all galaxies have fully calibrated optical images in DR16 \citep{SDSS-DR16}. However, the methodology outlined in \cite{Garcia-Benito:2015}, was designed for CALIFA observations, where galaxies mostly fill the entire PPak FoV. This facilitated a suitable match, enabling a relatively high number of fibers to capture galaxy light for aperture flux comparison. However, in the case of small and relatively faint galaxies, where only a few fibers are available for comparison, this methodology does not perform effectively. Indeed, for a few galaxies in the CALIFA survey with these characteristics, this registration approach did not work, as noted at the end of Sect. 4.2 in \cite{Garcia-Benito:2015}. This scenario mirrors what occurs with most of CAVITY galaxies, which occupy only a fraction of the FoV. In future iterations of the pipeline, we will explore adaptations of this methodology to tailor it to the specific characteristics of the CAVITY data and investigate the feasibility of employing a variant methodology utilising broad-band imaging registration in individual pointing frames. However, we use the SDSS broadband imaging for absolute flux calibration of the entire cube, as described later in this section.

After applying the Galactic extinction correction, the RSS is prepared for fiber spatial rearrangement and resampled spatially to form a data cube with a regular grid. We employed a flux-conserving inverse-distance weighting scheme \citep{Sanchez:2012}, which is a variation of Shepard’s interpolation method \citep{Shepard:1968}. The positions of each RSS pointing obtained in the preceding step are utilised for image reconstruction. We set the dispersion of the Gaussian to 0.75$\arcsec$ and constrain the kernel within a radius of 3.50$\arcsec$, resulting in a data cube with a spatial sampling of 1.00$\arcsec$, following \cite{Garcia-Benito:2015} and \cite{Sanchez:2016}.

The effect of differential atmospheric refraction (DAR) on data can be assessed empirically by tracing the centroid of the galaxy along the wavelength axis of the spatially resampled data cube. Typically, each monochromatic image slice is resampled to align the galaxy centroid to a common reference position. To streamline error propagation, a two-stage iteration is employed. First, the data cube is reconstructed, and the DAR effect is estimated. Next, the data cube is reconstructed again with fiber positions adjusted based on empirically measured DAR offsets. While error vectors are well-defined for each spectrum in reconstructed cubes, the inverse-distance weighting scheme introduces signal correlation in the spatial dimension, discussed in more detail in Sect. \ref{sec:noise}.

For optimal absolute flux calibration, considering the impracticality of individually registering pointing frames on a fiber basis, we conducted recalibration using large aperture SDSS photometry, as detailed in the methodology of \cite{Sanchez:2012} and \cite{Husemann:2012}, which was implemented in CALIFA DR1. The broad-band SDSS photometry is available for all targets as per the CAVITY sample design. Notably, the V500 grating data comprehensively covers the $g$ ($\lambda_{eff}$ = 4728 Å) and $r$ ($\lambda_{eff}$ = 6200 Å) filters \citep{Doi:2010} among the five SDSS filters ($ugriz$). To carry out the primary recalibration, we first measured the counts for each galaxy within a 30\arcsec\ diameter aperture in the SDSS images. These counts were then converted to flux density using the counts-to-magnitude formula provided in the SDSS documentation\footnote{\url{https://live-sdss4org-dr16.pantheonsite.io/algorithms/fluxcal}}. We then extracted spectrophotometric data from the reduced V500 data cubes. This involved summing the flux from individual spectra within a 30\arcsec\ diameter aperture and convolving the resulting spectrum with the SDSS $g$ and $r$ filter passbands. We proceeded to derive a scaling solution by averaging the flux ratios in both bands and adjusted the data cube flux and error accordingly.

Similar to the registration process at the individual pointing level, a second-order correction can be applied at the cube level to align the cube spectrophotometry with the corresponding broad-band images. The ratio between broad-band and reconstructed cube images can be computed to adjust fluxes across the data cube without altering spectral shape. The FLAT correction, first introduced in CALIFA DR3 \citep{Sanchez:2016} and subsequently applied in the eCALIFA remastered sample \citep{Sanchez:2023}, was incorporated into the CALIFA data cubes as an additional extension. However, this method is effective only when the galaxy covers a substantial portion of the cube and there are enough high S/N spaxels for the matching procedure. In the case of the CAVITY sample, similar to individual pointing registration, these conditions are not met for a significant portion of the targets, rendering the method inapplicable. In future pipeline versions, we will explore alternative methodologies tailored to the characteristics of the CAVITY dataset.

Finally, an astrometric registration of the data cube coordinate systems to the International Celestial Reference System (ICRS) is performed. This step is important when combining CAVITY data with other data, such as imaging or radio data, to extract spatially resolved information for specific regions in a galaxy. Since the PPak instrument does not have an built-in astrometric set-up, the CAVITY pipeline employs a straightforward approach: it uses tabulated coordinates from the \cite{Pan:2012}, which are assigned to the measured barycenter of the CAVITY data cubes. The CAVITY astrometry should be regarded as a general reference. Users are encouraged to visually inspect and adjust the World Coordinate System (WCS) information if precise spatial alignment is required for comparison with other datasets.

\subsection{Noise covariance \label{sec:noise}}

The interpolation procedure used to achieve a regular grid causes the output pixels in the final data cube to be interdependent. The Gaussian interpolation method distributes flux from a given fiber across multiple pixels, creating correlated noise between adjacent pixels. This correlation leads to an underestimation of noise in a stacked spectrum of N pixels, affecting statistical error estimates when coadding spectra in data cubes. Therefore, it is crucial to account for this noise covariance to ensure correct error propagation when combining or spatially binning spectra to enhance the S/N ratio, which is often necessary in analyzing IFS data.

One approach to addressing noise covariance is to compute the covariance matrix \citep{Sharp:2015}. In contrast, following the methodology of \cite{Garcia-Benito:2015}, we adopted an empirical approach to account for noise correlation by introducing the noise correction factor $\beta$(N). This factor is the ratio between the error estimated on the combined spectrum and the error derived by adding the individual errors of each spectrum quadratically.

To estimate this parameter, we randomly generated a set of circular apertures of various sizes (N spaxels) on each data cube, ensuring at least 20 apertures of the same size at different center positions on spaxels with a S/N $>$ 3. For each aperture, we empirically estimated the error of the coadded spectrum as the standard deviation from the detrended spectrum within the 5590–5710 \AA\ range in the rest frame. We also calculated the expected error, assuming no covariance, by propagating the individual empirical errors (measured in the same window and in the same way) within the aperture. The ratio between these two errors provides the $\beta$ value for the corresponding spectrum.

\begin{figure}
\includegraphics[width=\columnwidth]{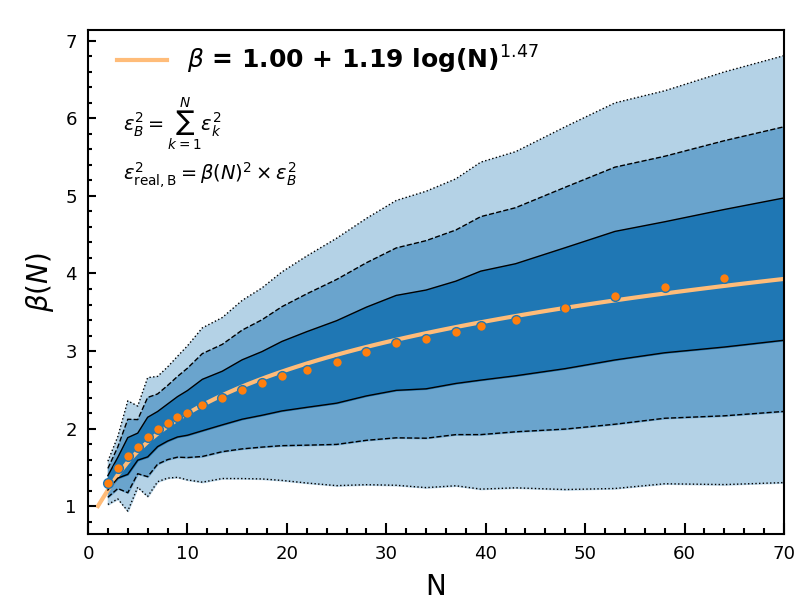}
 \caption{Correction factor to the noise introduced by the spatial covariance, $\beta$(N) (ratio of the real estimated error to the analytically propagated error), as a function of the number of co-added adjacent spaxels for all the DR1 V500 data. Shaded regions indicate the 1$\sigma$, 2$\sigma$, and 3$\sigma$ width areas around the average value, represented by solid circles. The solid line shows the best-fitted model, following logarithmic parameterisation, as shown in the legend.}
  \label{fig:beta}
\end{figure}

Figure \ref{fig:beta} illustrates the distribution resulting from over 100k estimations of $\beta$(N) on the CAVITY DR1 data as a function of the number of co-added spaxels/spectra (N). Following \citet{Garcia-Benito:2015}, we parameterised this distribution using a logarithmic function. However, unlike previous estimations, we included an exponent term to better fit the shape of the median distribution:

\begin{equation}
\beta(N)= 1 + 1.19 \log(N)^{1.47}
.\end{equation}

This function can then be applied to the noise spectrum from the data cube by multiplying the $\beta$(N) parameter by the error of $\epsilon_B$, the quadratic sum of the vector noise of each spectrum in the aperture.

\section{CAVITY data format\label{sec:format}}

The final reduced data for each observed object are consolidated into a single file, following the standard binary FITS format, and organised as data cubes, each comprising four FITS header/data units (HDU). The format mirrors that of CALIFA \citep{Husemann:2013} and other released IFS galaxy surveys. This multi-extension FITS file comprises various outcomes of the reduction process, detailed in Table \ref{tab:format}. These data cubes encompass (1) measured flux densities, corrected for Galactic extinction, measured in units of 10$^{-16}$ \uflux\ (primary data cube), (2) associated errors, (3) error weighting factors, (4) flags denoting bad pixels, and (5) fiber coverage. These data structures enable users to appropriately consider dataset characteristics such as bad pixels and noise for their specific analyses. 

The bad pixel FITS extension identifies pixels interpolated due to insufficient data from cosmic rays, defective CCD columns, or vignetting, and assigns them a large error value ($\sim$10$^{10}$) in the error vector stored in the error HDU extension. These bad pixels are unevenly distributed within the data cube, with four bad CCD columns reducing the number of valid pixels in specific wavelength regions. The hexagonal PPak FoV is resampled to a rectangular grid, where uncovered corners are flagged as bad pixels, while residuals from bright night-sky emission lines remain unflagged.

The spatial dimensions of the data cubes are delineated along right ascension and declination in the first two axes, each sampled at 1\arcsec $\times$ 1\arcsec. The third axis represents the wavelength, sampled on a linear grid with 2 \AA\ steps. Table \ref{tab:cube_dimension} provides a summary of the dimensions of each data cube (N$\alpha$, N$\delta$, and N$\lambda$), as well as the spectral sampling (d$\lambda$) and resolution ($\delta_{\lambda}$) across the entire wavelength range. 

These files are named after the official object names from the CAVITY mother sample, adopting the format CUBENAME.V500.rscube.fits, with CUBENAME corresponding to the OBJECT header. 

\begin{table*}
\centering
\caption{CAVITY FITS file structure.}
%\label{tab:HDUs}
\label{tab:format}
\begin{tabular}{cccc}\hline\hline
HDU & Extension name & Format & Content\\\hline
0 & Primary & 32-bit float & Flux density in units of $10^{-16}\,\mathrm{erg}\,\mathrm{s}^{-1}\,\mathrm{cm}^{-2}\,\mathrm{\AA}^{-1}$\\
1 & ERROR & 32-bit float & $1\sigma$ error on the flux density\\
2 & ERRWEIGHT & 32-bit float & Error weighting factor\\
3 & BADPIX &   8-bit integer & Bad pixel flags (1=bad, 0=good) \\
4 & FIBCOVER &   8-bit integer & Number of fibers used to fill each spaxel \\
\end{tabular}
\end{table*}

\begin{table}
\caption{Dimension and sampling of CAVITY data cubes.}
\label{tab:cube_dimension}
\begin{tabular}{lccccccc}\hline\hline
\small{Set-up} & \small{$N_\alpha$}\tablefootmark{a} & \small{$N_\delta$}\tablefootmark{a} & \small{$N_\lambda$}\tablefootmark{a} & \small{$\lambda_\mathrm{start}$}\tablefootmark{b} & \small{$\lambda_\mathrm{end}$}\tablefootmark{c} & \small{$d_\lambda$}\tablefootmark{d} & \small{$\delta_\lambda$}\tablefootmark{e} \\\hline
\small{V500}  &  \small{78}    & \small{73}    &  \small{1877}   & \small{3749\AA}     & \small{7501\AA}  & \small{2.0\AA}  & \small{6.0\AA}    \\
\end{tabular}
\tablefoot{\tablefoottext{a}{Number of pixels in each dimension.}
\tablefoottext{b}{Wavelength of the first pixel on the wavelength direction.}
\tablefoottext{c}{Wavelength of the last pixel on the wavelength direction.}
\tablefoottext{d}{Wavelength sampling per pixel.}
\tablefoottext{e}{Homogenised spectral resolution (FWHM) over the entire wavelength range.}
}
\end{table}

%\subsection{FITS header information}
The FITS header encompasses the standard keywords that map the spatial axes onto the standard World Coordinate System \citep{Greisen:2002}, and the wavelength onto the spectral axis within a linear grid. Within each CAVITY data cube resides the complete FITS header information of all constituent raw frames, wherein each header entry is augmented with a distinct prefix corresponding to a specific pointing or frame. This prefix comprises the grating utilised (\texttt{GRAT}; V500 in this DR), followed by `PPAK' and the identification of the pointing frame $\tt P_{i}F_{j}$, where $P_i$ denotes the pointing (ranging from 1 to 3), and $F_j$ represents the frame within a particular $P_i$ pointing (with $j$ varying from 1 to 4 based on the luminosity of the galaxy; two frames for bright galaxies and four frames for low-luminosity galaxies). Information pertaining to the combined pointing, where all individual frames are amalgamated into a single pointing file, follows the structure of \texttt{GRAT} PPAK $P_i$. Moreover, the reduction pipeline aggregates supplementary details such as sky brightness, Galactic extinction, etc., and integrates them into the FITS header. Header keywords that hold potential significance for data analysis and/or assessment are succinctly outlined in Table \ref{tab:header_keys} for ease of reference.

\begin{table*}
\DeclareDocumentCommand{\ijkw}{ m }{hierarch \texttt{GRAT} PPAK $\tt P_{i}F_{j}$ #1}
\DeclareDocumentCommand{\ikw}{ m }{hierarch \texttt{GRAT} PIPE $\tt P_{i}$ #1}
 \caption{Main FITS header keywords}
 \label{tab:header_keys}
 \centering
 \begin{tabular}{lll}\hline\hline
  Keyword & Data type & Meaning\\\hline
  OBJECT   & string  & Name of the target galaxy (CAVITY ID)\\
  NAXIS1   & integer & Number of pixels along right ascension axis  ($N_\alpha$)\\
  NAXIS2   & integer & Number of pixels along declination axis  ($N_\delta$)\\
  NAXIS3   & integer & Number of pixels along wavelength axis  ($N_\lambda$)\\
  CRPIX1   & float   & Reference pixel of the galaxy center in right ascension\\
  CRVAL1   & float   & Right ascension $\alpha$ (J2000) of the galaxy center in degrees\\
  CDELT1   & float   & Sampling along right ascension axis in arcsec\\
  CRPIX2   & float   & Reference pixel of the galaxy center in declination\\
  CRVAL2   & float   & Declination $\delta$ (J2000) of the galaxy center in degrees\\
  CDELT2   & float   & Sampling along declination axis in arcsec\\
  CRVAL3   & float   & Wavelength of the first pixel along the wavelength axis in \AA \\
  CDELT3   & float   & Sampling of the wavelength axis in \AA\\
  hierarch CAVITYVER & string & Version of the CAVITY reduction pipeline\\
  hierarch PIPE UNITS & string & Units of the flux density\\
  hierarch PIPE GALEXT COR & integer & Applied Galactic Extinction (0 No, 1 Yes) \\
  hierarch PIPE GALEXT AV & float & Galactic $V$ band extinction along line-of-sight\\
  hierarch PIPE REDUDATE & string & Date/time the data were reduced\\
  \ijkw{DATE-OBS} & string & Date/time of the observation\\
  \ijkw{UT\_START} & float & Start of the observation in seconds UT time\\
  \ijkw{UT\_END} & float & End of the observation in seconds UT time\\
  %\ijkw{MJD-OBS} & float & Modified Julian date of the observation\\
  \ijkw{AIRMASS} & float & Airmass during the observation\\
  \ijkw{EXPTIME} & float & Exposure time of the observation\\
  \ijkw{HVCOR} & float & Line-of-sight difference to heliocentric velocity\\
  \ijkw{EXT\_V} & float & $V$ band atmospheric extinction in magnitudes\\
  %hierarch PPAK \textit{PREF} PIPE FLEX XOFF & float & Measured flexure offset in $x$-direction in CCD pixels\\
  %hierarch PPAK \textit{PREF} PIPE FLEX YOFF & float & Measured flexure offset in $y$-direction in CCD pixels\\
  \ijkw{PIPE SPEC RES} & float & Homogenised spectral resolution (FWHM) in \AA \\
  \ijkw{PIPE NSKY FIB} & integer & Number of averaged sky fibers \\ 
  \ijkw{PIPE SKY MEAN} & float & V Johnson mean surface brightness ($\mu_\mathrm{sky}$) during observation \\
  \ikw{RELFLUX RATIO MEDIAN} & float & Absolute flux scaling of the combined pointing\\
  \hline
 \end{tabular}
\end{table*}

\section{Data quality\label{sec:qc}}

In order to select the 100 galaxies to be released in the CAVITY DR1, we have run a careful quality control (QC) on the resulting data cubes to ensure that the released data comply with the standard requirements for their use by the scientific community. In this section we detail the QC checks that have been performed for such purpose. These checks are first based on a visual inspection of the data cubes by the QC team (Sect.~\ref{sec:qc_visual}), which is followed by a final quality assessment based on a set of measured parameters (Sect.~\ref{sec:qc_auto}).

\subsection{Visual quality checks\label{sec:qc_visual}}

A visual inspection of each final reduced data cube is performed to estimate the quality of the data and identify any problem with the observations or the reduction that highly affects the cube reconstruction and/or the individual spectra. This QC is carried out based on a series of graphics that are automatically generated for each data cube after one round of observations is reduced. To quickly assess the quality of the data and their suitability for scientific use, four flags were defined:

\begin{enumerate}
    \item \textsc{cube}. This flag allows us to identify problems related to the cube reconstruction, such as the presence of a grid associated with the dithering scheme (which might uncover issues with a particular pointing) or a target with very low surface brightness unsuitable for our observing strategy. To flag these problems, a white image of the galaxy (from 4500 to 7000 \AA) is provided together with a map of its negative values (low-transmission fibers clearly stand out in this plot, see Sect.~\ref{sec:deadfibers}).\\[-0.2cm]
    
    \item \textsc{spectra}. We indicate here if the spectra present any distortion or odd features, or an atypical shape in the continuum emission. For that, the reviewers are given for each galaxy the central spectrum, together with the integrated spectrum within 0.5, 1, and 2 effective radii.\\[-0.2cm]
    
    \item \textsc{vignetting}. Overall, CAVITY galaxies tend to be small ($d_{25}<$ 40 arcsec), so the vignetting effect (concentrated in an annular ring at approximately $\sim$15\arcsec\ from the center of the FoV, see Sect.~\ref{sec:obs} for more details) only affects the observations in the very outer parts of the galaxies. However, for larger galaxies, or if there is an offset of its center with respect to the center of the FoV, this effect can impact the spectra of the inner parts and hamper for instance the analysis of the stellar populations. This flag allows us to identify such cases, by means of observing a clear jump at the ends of the integrated spectra within 0.5 or 1 effective radii or in the white image if the galaxy extends across the vignetting area marked with a ring of semi-transparent dots (a contour indicating the 23 mag/arcsec$^2$ level of the galaxy surface brightness allows us to delimit the galaxies). \\[-0.2cm]
    
    \item \textsc{jumpccd}. The reported issue with the CCD quadrant `a' (see Sect.~\ref{sec:ccd} for details) produced for some observations the presence of a jump in the spectra at around 5800 \AA\ and a sharp step in the sky level between the top and bottom halves of the instrument FoV (see also Sect.~\ref{sec:grad}). All observations affected by this issue were discarded and the galaxies were re-observed once the problem was fixed. This flag helps to identify these cases after inspecting (i) a zoom-in image of the integrated spectra within 0.5, 1, and 2 effective radii in the spectral range covering from 5500 to 6100 \AA\ (where the spectral jump stands out), and (ii) an image displaying the average flux in each row of pixels when scanning the FoV from north (top) to south (bottom) after removing the central region where the galaxy is located (where the spatial jump is clearly observed).
\end{enumerate}
In addition, any other  issue not covered by these flags was reported as a supplementary comment. The visual assessment of each data cube is done by two different members of the team. When any of the described flags are marked as \textsc{bad} by the two members, we consider the data cube unsuitable, and the galaxy is discarded from the DR1 candidate sample. When the two reviewers disagreed on any of the flags, the examination of a third member of the QC team breaks the discrepancy and decides whether the galaxy passes the QC visual inspection and enters the set of candidates from which the DR1 sample will be drawn or is discarded from further QC checks. All the galaxies with \textsc{good} values of the four described flags pass to the second stage of the QC assessment, detailed in the next subsection.

\subsection{Automatic quality checks\label{sec:qc_auto}}
This second phase of the QC assessment focuses on inspecting a set of measured parameters, extracted by the pipeline at different stages of the reduction process, which will help to characterise different aspects in the quality of the released data cubes. These parameters are categorised in three groups: those related to the observing conditions (denoted by the \textsc{obs} prefix), those associated with the instrumental configuration and the performance of the data reduction pipeline (\textsc{red}), and those evaluating the accuracy and quality of the final calibrated data products (\textsc{cal}). For the assessment, different thresholds are defined for each parameter mainly following the procedure adopted for the CALIFA DR3. This decision is based on the use of the same instrument by both CALIFA and CAVITY projects, and similar observing strategies and reduction pipelines. The thresholds were first determined based on the distributions of the parameters (to flag obvious outliers). Along the CALIFA DRs, these were modified based on the larger statistics available and larger experience with the data, checking on the impact of exceeding such thresholds on the final quality of the data. Below we describe the explored parameters and the corresponding flags, but we encourage the reader to consult \citet{Sanchez:2016} for more detailed information. Since the parameters are measured normally on the individual pointings and/or fibers (depending on the parameter), the thresholds are applied to different measurements obtained when combining the individual estimations. The defined flags will involve in general the mean value (mean), the maximum value (max), and the standard deviation $\sigma$ (std) of the different estimations of each quantity. For convenience, the used flags and their corresponding thresholds are summarised in Table~\ref{tab:QC}.  The values of these flags for the entire DR1 sample are provided on the project webpage\footnote{\urls{\urltabs}}. From the original sample of 246 galaxies observed up to June 2024, we remove those with \textsc{BAD} flags in any of the analysed parameters. It is from this subset that the final 100 DR1 galaxies are randomly selected, ensuring a proper coverage of the color-magnitude diagram below -18 mag in $r$-band (see Fig.~\ref{fig:sample_CMD} and related information provided in Sect.~\ref{sec:sample}). 

\begin{table*}
\centering
\caption{CAVITY DR1 defined quality control flags and thresholds for the automatic checks described in Sect.~\ref{sec:qc_auto}.}
\label{tab:QC}
\begin{tabular}{llll}\hline\hline\\[-0.3cm]
Parameter involved\hspace{2cm} & QC flags\hspace{3cm} & \textsc{WARNING} condition & \textsc{BAD} condition  \\\hline\\[-0.2cm]
Airmass & \textsc{obs\_airmass\_mean} & $>2$ & -- \\
 & \textsc{obs\_airmass\_max} & $>2.5$ & -- \\
 & \textsc{obs\_airmass\_std} & $>0.15$ & -- \\[0.2cm]
Sky brightness & \textsc{obs\_skymag\_mean} & $<19.5$ mag arcsec$^{-2}$ & -- \\
 & \textsc{obs\_skymag\_std} & $>0.1$ & -- \\[0.2cm]
Atmospheric extinction & \textsc{obs\_ext\_mean} & $>0.3$ mag & -- \\
 & \textsc{obs\_ext\_max} & $>0.35$ & -- \\
 & \textsc{obs\_ext\_std} & $>0.1$ & -- \\[0.2cm]
Straylight & \textsc{red\_meanstraylight\_max} & $>50$ counts & $>100$ counts \\
 & \textsc{red\_maxstraylight\_max} & $>75$ & $>150$ \\
 & \textsc{red\_stdstraylight\_max} & $>15$ & $>30$ \\[0.2cm]
Spectral dispersion & \textsc{red\_disp\_mean} & $>5.5$ \AA\ (FWHM) & $>10$ \\
 & \textsc{red\_disp\_max} & $>10$ & -- \\
 & \textsc{red\_disp\_std} & $>0.5$ & -- \\[0.2cm]
Cross dispersion & \textsc{red\_cdisp\_mean} & $>3$ pixels (FWHM) & -- \\
 & \textsc{red\_cdisp\_max} & $>4$ & -- \\
 & \textsc{red\_cdisp\_std} & $>0.25$ & -- \\[0.2cm]
Sky subtraction & \textsc{red\_res5577\_min} & $<-0.1$ counts & -- \\
 & \textsc{red\_res5577\_max} & $>0.1$ & -- \\
 & \textsc{red\_stdres5577\_max} & $>1.0$ & -- \\[0.2cm]
Spectrophotometric calibration & \textsc{cal\_specphot\_fluxratio} & -- & $<0.7$  \\
& &  & $>1.6$ \\\hline
\end{tabular}
\end{table*}

\subsubsection{Quality of the observing conditions}
To evaluate the quality of the observing conditions we focus on three main quantities: the airmass, the sky brightness, and the atmospheric extinction. As explained in CALIFA DR3, observing conditions alone do not qualify to consider a data cube to be unsuitable for science, but they raise a warning when the defined thresholds are exceeded; this indicates the presence of minor issues slightly affecting the quality of the data.

For the airmass, we consider three different flags based on the average (\textsc{obs\_airmass\_mean}), the maximum (\textsc{obs\_airmass\_max}), and the $\sigma$ (\textsc{obs\_airmass\_std}) values over all contributing pointings. When planning the observations, we request a minimum elevation of $\sim45^{\circ}$ for the targets, which corresponds to an airmass of $1.4$ (well below the set warning threshold). This implies that only the \textsc{std} flag can raise a warning when its threshold is exceeded, which happens in only 3\% of the cases.

The sky brightness is measured in the V-band in each pointing using the 30 faintest sky fibers (see Sect.~\ref{sec:obs}). In this case, the mean (\textsc{obs\_skymag\_mean}) and the $\sigma$ (\textsc{obs\_skymag\_std}) values over all pointings are considered to define the corresponding flags. The threshold of 19.5 mag arcsec$^{-2}$ for the average is never passed; however, the limit of 0.1 mag arcsec$^{-2}$ for $\sigma$ is exceeded in 36 galaxies. This is presumably related to the sky gradient that is observed in several galaxies (more details can be found in Sect.~\ref{sec:grad}), as the average value of the sky fibers is used to define and subtract the sky level (see Sect.~\ref{sec:sky}).

Finally, the V-band extinction at the time of each observed pointing is monitored by the observatory and allows us to characterise the transparency of the sky. We provide thresholds for the mean (\textsc{obs\_ext\_mean}), maximum (\textsc{obs\_ext\_max}) and $\sigma$ (\textsc{obs\_ext\_std}) values across the pointings (reported in Table~\ref{tab:QC}). A warning is raised in 13\% of the cases, indicating inhomogeneous observing conditions.

\subsubsection{Quality of the instrumental/data reduction performance}

The performance of the instrument and the data reduction is assessed based on four properties: straylight, spectral dispersion, cross dispersion, and residuals from the subtraction of bright skylines. These quantities are measured on the reduced data (individual fiber spectra), before combining them to create the final data cubes. 

The straylight (see Sect.~\ref{sec:spext}) is subtracted from the spectra by the reduction pipeline, but high residuals can affect the final quality of the data. To ensure that this is not the case for the CAVITY DR1 galaxies, we set some thresholds on the mean (\textsc{red\_meanstraylight\_max}), maximum (\textsc{red\_maxstraylight\_max}), and $\sigma$ (\textsc{red\_stdstraylight\_max}) values for all the contributing frames (and thus the \textsc{\_max} suffix). No warnings in the DR1 sample have been raised due to low levels of these quantities.

The spectral dispersion (\textsc{disp}) and cross dispersion (\textsc{cdisp}) are measured on the arc-lamp calibration frames and the continuum lamp data as the FWHM of the line and the FWHM of the spectral trace, respectively (Sect.~\ref{sec:spext}). Thresholds are set on the mean (\textsc{red\_disp\_mean}, \textsc{red\_cdisp\_mean}), maximum (\textsc{red\_disp\_max}, \textsc{red\_cdisp\_max}), and $\sigma$ (\textsc{red\_disp\_std}, \textsc{red\_cdisp\_std}) values to avoid significant departures of both parameters from the nominal target specifications. 12 and 7 galaxies of the sample raise a warning because of the requirements on the spectral dispersion and cross dispersion values, respectively.

The accuracy of the sky subtraction is quantified by measuring the flux residuals of the 5577~\AA\ \io{o}{i} skyline. We define the flags based on the minimum (\textsc{red\_res5577\_min}) and the maximum (\textsc{red\_res5577\_max}) values over all pointings of the fiber-average residuals, and the maximum (over all pointings) of the $\sigma$ of the residuals (\textsc{red\_stdres5577\_max}). As explained in \citet{Sanchez:2016}, average residuals that are too negative or too positive are indicative of systematic bias in the sky subtraction, whereas large $\sigma$ can reflect localised failures or noisy data. Large enough residuals and/or $\sigma$ appear in 24\% of the cases as to raise a warning, but in any case they are below $\sim0.5$ and $\sim1.5$ counts, respectively.

\subsubsection{Quality of the calibrated data products\label{sec:specphot}}

The quality of the calibrated data products is determined by checking the global spectrophotometry and the stability of the wavelength calibration across the spectral range.

As described in Sect. \ref{sec:cube}, each V500 data cube in the CAVITY pipeline was rescaled to match the absolute flux level of the SDSS DR16 broad-band photometry within a 30\arcsec\ diameter aperture. This flux rescaling process for absolute re-calibration is not perfect, as the V500 is slightly affected by vignetting in some fibers (see Sect. \ref{sec:obs}). Additionally, artifacts and minor contamination from field stars, which vary due to seeing conditions and PSF differences between the DR16 images and the data cube, also may affect slightly the flux within the aperture in different ways. We constrained the median scale factors (\textsc{cal\_specphot\_fluxratio}) to be between 0.7 and 1.6, removing objects with values outside these limits. Such outliers were primarily due to the partial coverage of the circular aperture in the DR16 images. 

\begin{figure*}
 \includegraphics[width=0.5\textwidth]{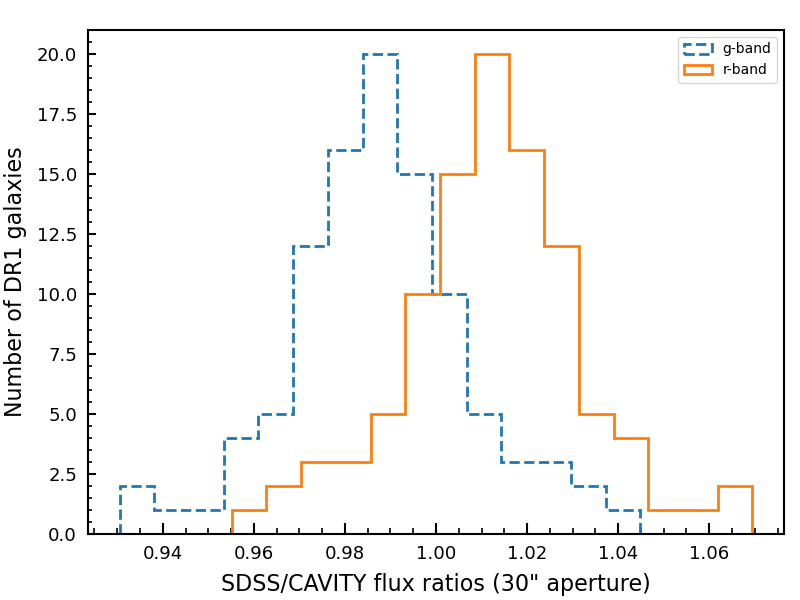}
 \includegraphics[width=0.5\textwidth]{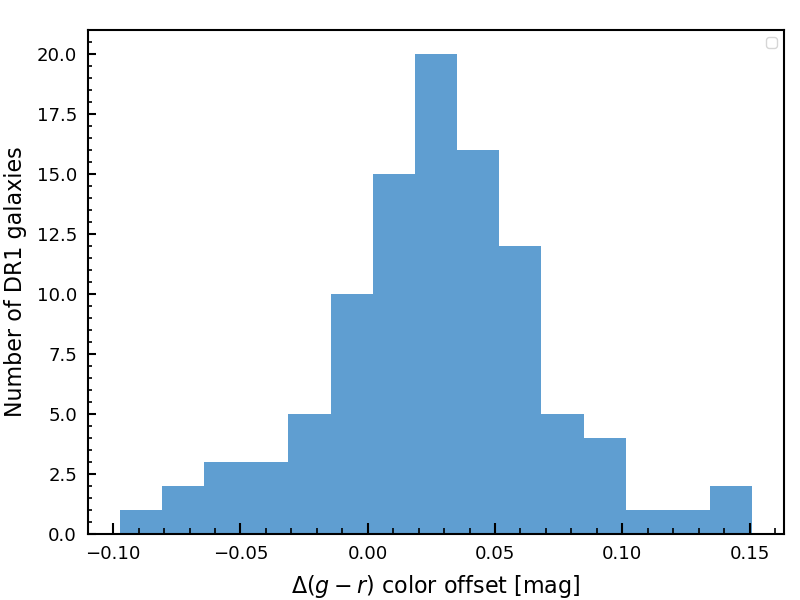}

 \caption{Distribution of the scale factor for 30\arcsec\ aperture photometry between the SDSS DR16 images and the recalibrated CAVITY data (left). Distribution of the color offset between the SDSS DR16 images and the synthetic broadband images derived from cubes in CAVITY (right).}
  \label{fig:flux_match}
\end{figure*}

The distribution of photometric scale factors for the $g$ and $r$ bands is shown in Fig. \ref{fig:flux_match} (left panel). The median SDSS/CAVITY $g$ and $r$ band ratios after recalibration are 0.99 $\pm$ 0.02 and 1.01 $\pm$ 0.02, respectively, indicating that the absolute photometric calibration of the CAVITY data is accurate to within 5\%. The right panel of Fig. \ref{fig:flux_match} shows the distribution of the $\Delta$($g$ - $r$) color difference between the SDSS and CAVITY data. We observe a slight systematic offset of $\Delta$($g$ - $r$) = 0.03 mag (median) with a scatter of only 0.04 mag. Thus, the spectrophotometric accuracy across most of the covered wavelength range is $\sim$ 5\% for the CAVITY data.

Finally, in order to evaluate the stability of the wavelength calibration over the full spectral range, we look at the position of several sky lines, comparing the nominal wavelength with the measured value in the observed spectra. In particular, we check the median, mean and $\sigma$ values across the FoV. In all cases, we find that the centroids are fully consistent with zero offsets from the nominal wavelength, and the $\sigma$ is consistent with pure measurement errors and the absence of any detectable systematic spatial variation.

\subsection{Data cube cosmetics\label{sec:qc_cosme}}

\begin{figure*}
 \includegraphics[width=0.95\textwidth]{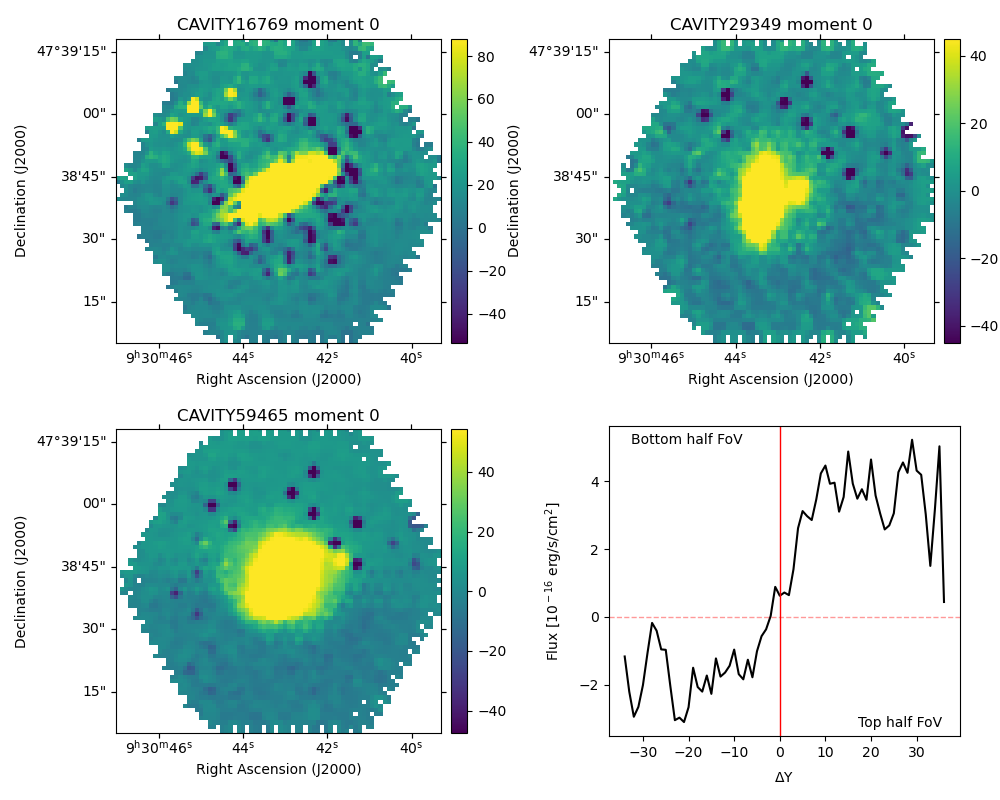} \\
\caption{Examples of CAVITY DR1 data cubes affected by cosmetic issues. \emph{Top-left :} CAVITY~16769 displays clear examples of bright and faint patches due to variable fiber transmission. \emph{Top-right:} CAVITY~29349 shows a patchy, hexagonal pattern in the central regions consequence of the dithering scheme. \emph{Bottom panels} exemplify the gradient in the faintest regions for CAVITY~59465. In the left we show the moment 0 image while the right-hand panel displays bottom-to-top profile of the background flux. All moment 0 images (from 4500 to 7000 \AA) are in units of 10$^{-16}$·erg·cm$^{-2}$·s$^{-1}$.}
  \label{fig:cosmetics}
\end{figure*}

Some of the CAVITY data cubes present aesthetic features that we describe and characterise in detail in this section. These include: (i) the presence of some fibers with variable transmission that are reflected in the reconstructed data cubes as bright or faint patches; (ii) a slight gradient visible in the lowest surface brightness regions (normally oriented north to south); and (iii) a hexagonal pattern in the central parts of the cubes of the smallest galaxies. Figure~\ref{fig:cosmetics} gives some examples of all cosmetic issues detected in CAVITY DR1 data cubes.

Before proceeding with the description and characterisation of all these features we must assure the reader about the effect of these features on the science to be carried out. During the CAVITY survey, larger galaxies similar to those targeted by CALIFA were also observed by other PMAS programs involving collaboration members and reduced using the CAVITY pipeline described in this work. The reconstructed cubes from these larger galaxies do not exhibit the cosmetic effects described in this section. This indicates that these effects are only noticeable at very low surface brightness in small galaxies, where a significant portion of the FoV is dominated by sky emission. Thus, although striking to the eye, these data cube cosmetics are mainly low-order features that do not affect the main science to be carried out with the data cubes, with the exception of variable transmission fibers that should be masked out (see the next section).

\subsubsection{Variable transmission fibers}\label{sec:deadfibers}

As we can observe from Fig.~\ref{fig:cosmetics}, CAVITY DR1 galaxies present some regions with very low or high transmission that are reflected in the reconstructed data cubes as faint or bright patches. Each fiber affected with poor transmission (high and low in the case of CAVITY~16769 or low for the rest), is repeated three times following the dithering scheme used for the observations (see Sect.~\ref{sec:obs}). These three examples highlight the variability of the phenomenon (not every galaxy display the same distribution of ill-behaved fibers).

\begin{figure}
    \includegraphics[width=\columnwidth]{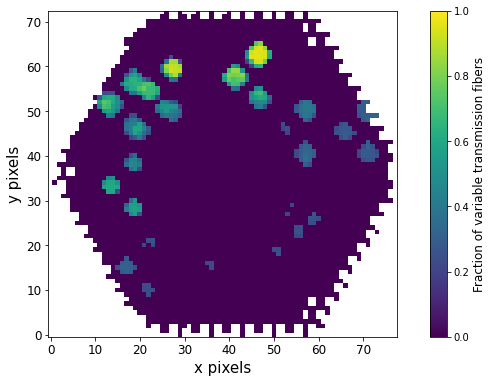}
    \caption{Fraction of data cubes in which a given pixel presents variable transmission along the FoV of the reconstructed data cubes. Only those present in more than 20$\%$ of galaxies are shown.}
    \label{fig:qc_dead_fib}
\end{figure}

Figure~\ref{fig:qc_dead_fib} displays the ubiquity of this issue as the fraction of galaxies for which a given pixel is affected by a fiber with variable transmission. For this, first we inspect, in a galaxy-by-galaxy basis, the presence and location of such fibers. Then we evaluate, for each pixel in the data cube, the number of galaxies out of the 100 CAVITY DR1 galaxies for which such pixel is affected by a transmission variance. Given the variable nature of the problem, this figure should not be interpreted as an effective mask to deal with this phenomenon, but rather as a statistical description of where, and how often, ill-behaved fibers are found in CAVITY data cubes. Plans exist for future reduction pipeline versions and future data releases to minimise this feature via the creation of masterflat fields tailored to specific observing epochs.

\subsubsection{Overall gradient}\label{sec:grad}

The relatively small sizes of the CAVITY galaxies imply that a good percentage of the PMAS/PPak area is not filled with the galaxy. For a small number of systems, these areas, mainly dominated by sky emission, display a low-amplitude background gradient. This gradient is usually oriented from north to south (top to bottom), although some exceptions are found. 

It displays a north-to-south median flux profile of the data cube (vertical scan of all pixels after the galaxy is removed, see Sect.~\ref{sec:qc_visual}, point {\it 4}). In general, the low amplitude of these gradients have a negligible effect in the regions where we detect light from the observed galaxies (a factor of $\gtrsim$100 brighter), however, it is of the order of the flux in the sky-dominated areas after sky subtraction. Thus, although visible in a few galaxies, it does not affect the regions of scientific interest.

\begin{figure}
    \resizebox{\hsize}{!}{\includegraphics[trim=1.6cm 0 2.8cm 2cm,clip]{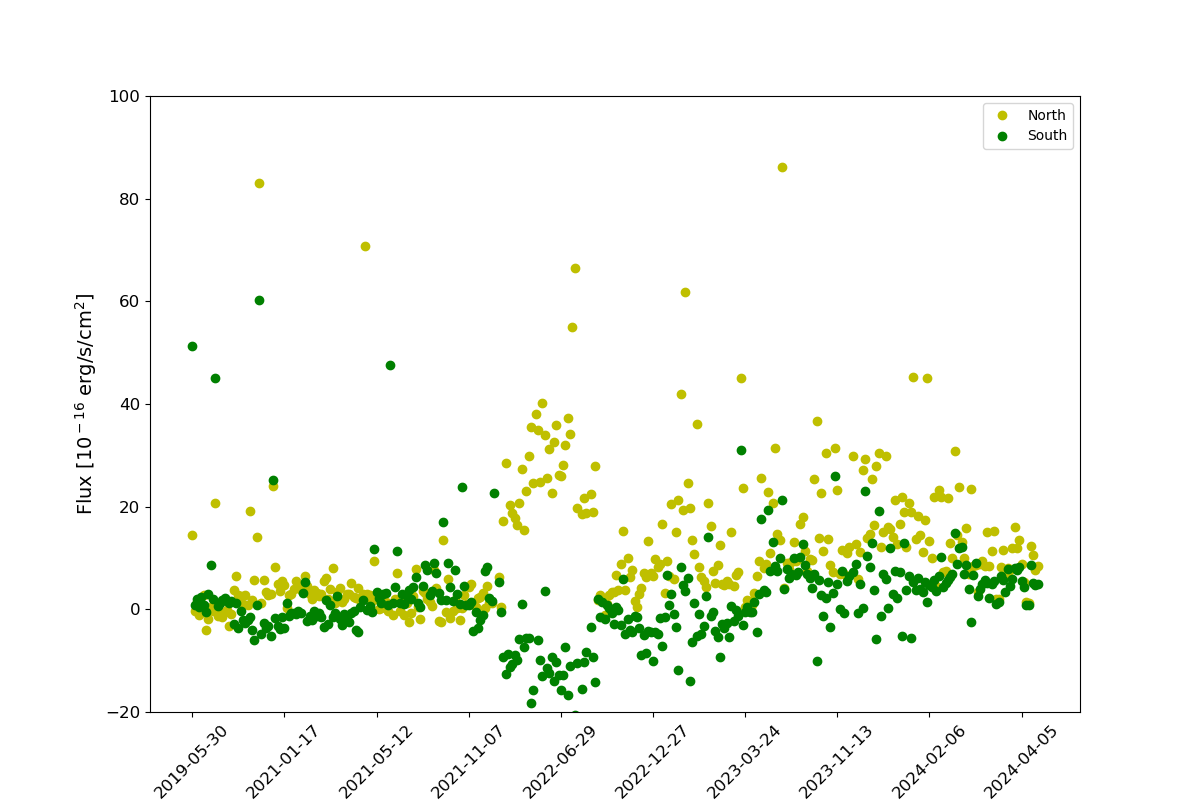}}
    \caption{Time evolution of the median flux in background regions located towards the north (yellow) and south (green) of the CAVITY DR1 data cubes (avoiding spaxels with contribution from the observed galaxy).}
    \label{fig:grad_time}
\end{figure}

To fully assess the stability of this feature, we analyze its evolution over time in Fig.~\ref{fig:grad_time}. For doing that, we measure the median flux in a northern (yellow) and southern (green) background area (avoiding light from the galaxy) from data cubes observed during the different observing campaigns (x-axis). We highlight the stability of the PMAS observations from 2019 to December 2021 (low difference in flux between north and south). It is then when a clear difference between the northern and southern areas starts to appear (signs of a high amplitude gradient), coinciding with the CCD failure reported before. From November 2022 onward (official resumption of the CAVITY observations) this difference was hugely minimised, with some epochs of slightly larger gradients coinciding with winter observations.

From this analysis we can conclude that most of the gradients (and all of the steep gradients) detected during the assessment of the quality of the CAVITY data are associated with the CCD failure reported before. These data cubes will not be part of any CAVITY data release as they do not fulfill the minimum quality requirements. Also, we would like to highlight that galaxies observed during this period have been re-observed over the past year to ensure they were observed under optimal observing conditions. As a consequence, only a handful of CAVITY DR1 galaxies present low amplitude gradients. 

\subsubsection{Central low-surface brightness pattern}\label{sec:dit_cent}

Another low-order cosmetic artifact observed in a minority of DR1 CAVITY data cubes is a hexagonal-like patchy pattern, exemplified by CAVITY~29349 (see Fig.~\ref{fig:cosmetics}). This pattern results from the dithering scheme used during observations and the slight flux variations between pointings. Although all pointings are flux-matched as described in Sect. \ref{sec:cube}, these subtle differences become more apparent when the surface brightness is low. This artifact is seen only in the central regions of the lowest surface brightness, underscoring its low amplitude. Consequently, it is mainly observed in the smallest CAVITY galaxies.

\section{Access to the CAVITY DR1 data\label{sec:access}}

CAVITY DR1 comprises cubes and catalogs containing essential variables and derived physical properties for 100 CAVITY galaxies within 15 voids. All the cubes and catalogs of this public data release are available on the survey’s dedicated web page\footnote{\urls{cavity.caha.es}}. This  web page serves as the primary hub for the survey, where all updates, news, and future data releases related to both CAVITY and CAVITY+ will be provided \citep[for more information on CAVITY+ see][]{Perez:2024}. Constructed using the Open Source Daiquiri framework \citep{2020Galkin}\footnote{\url{https://github.com/aipescience/django-
daiquiri}}, the data release web page facilitates SQL query storage, blog notifications, and commenting functionality, with optional login features. For convenience, all CAVITY catalogs are provided in FITS and CSV formats. The tables are also compatible with the Virtual Observatory (VO) and are accessible using Table Access Protocol\footnote{\urls{cavity.caha.es/tap}} \citep[TAP;][]{2011ASPC..442..603D}. In Fig.~\ref{Landing_page}, a screenshot captured in July 2024 showcases the project home page. A brief description of each section of the webpage is provided. In what follows, we explain the properties and information contained in each part of the web page.

\subsection{Project \& data tabs} \label{sec:access1}
In the "project" tab, users can find general information about the project and CAVITY team members. The "data" tab provides news and details about CAVITY data releases. As outlined in \cite{Perez:2024}, the CAVITY project extends to different wavelengths by taking advantage of available data from public surveys and conducting dedicated observing campaigns. Value-added catalogs, products of various projects performed by the team members over the CAVITY and CAVITY+ datasets, are published and accessible through this tab.

\subsection{Database tables}\label{sec:access2}
The metadata description for DR1 can be found under the "database tables" drop-down menu on the "CAVITY DR1" page. This page includes a list of table icons that provide various information and parameters for CAVITY galaxies. The "PPAK FITS files" table contains data cubes of the released CAVITY galaxies. Users can access the footprint and radar imagery of the CAVITY galaxies in the "PPAK-images" table. These images show the spatial coverage of the PPAK hexagonal footprint relative to the galaxy's effective radius. The other tables on this page include additional related parameters for CAVITY galaxies, gathered through cross-correlation with other large multi-wavelength surveys. Detailed descriptions of all tables on the "CAVITY DR1" page are presented on the project webpage\footnote{\urls{\urltabs}}.  

In each sub-page of tables, users can find comprehensive information about the columns of tables and references for the data sources. Each page starts with a short introduction to the table and its references, followed by a table listing all the physical parameters available in unified content descriptors\footnote{\url{https://ivoa.net/documents/UCD1+/}} \citep[UCDs, from the International Virtual Observatory Alliance;][]{2004Derriere}, and brief descriptions. These catalogs can be accessed via SQL query in both CSV and FITS formats (see Sect. \ref{sec:access3} for an example).

The "CAVITY master catalog" serves as a comprehensive summary of the main physical properties of the galaxies, extracted from all the catalogs accessible on the "CAVITY DR1" page. As such, it stands as an excellent initial resource for users aiming to get familiar with the sample and pursue additional scientific inquiries. Specifics of the "master" table are outlined in Table~\ref{tab:CAVITY_master_table}. 

\begin{figure*}
    \centering
    \includegraphics[scale = 0.5]{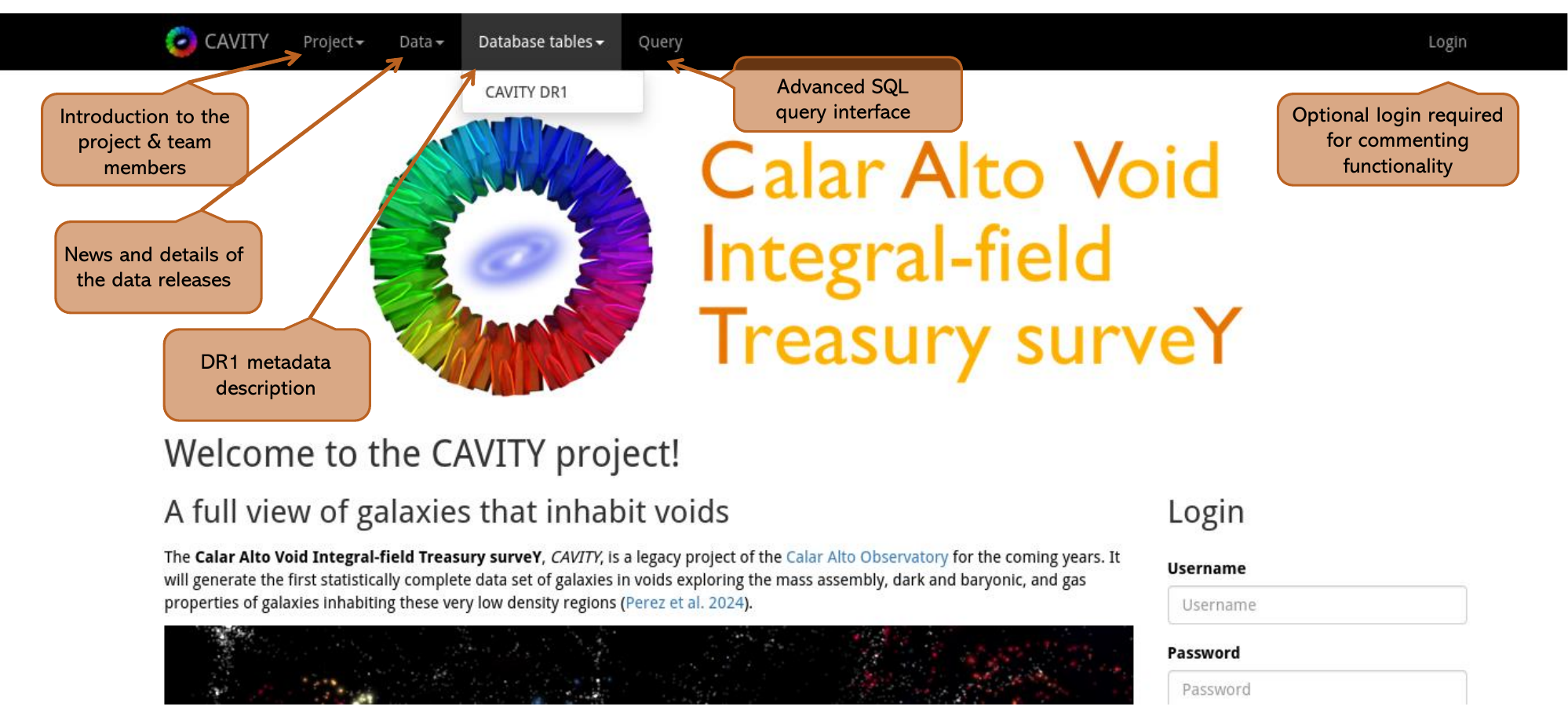}
    \caption{Screenshot of the landing page of CAVITY website captured in July 2024. CAVITY data and catalogues are accessible through various tabs and an advanced SQL query interface. }
    \label{Landing_page}
\end{figure*}

\begin{table*}
\DeclareDocumentCommand{\ijkw}{ m }{hierarch \texttt{GRAT} PPAK $\tt P_{i}F_{j}$ #1}
\DeclareDocumentCommand{\ikw}{ m }{hierarch \texttt{GRAT} PIPE $\tt P_{i}$ #1}
 \caption{The CAVITY master table$^{*}$}
 \label{tab:CAVITY_master_table}
 \centering
 \begin{tabular}{lll}\hline\hline
  Physical parameter & data type & Description\\\hline
  galaxy   & integer  & Name of the target galaxy (CAVITY ID)$^{a}$\\
  void   & integer & The hosting void number$^{a}$\\
  ra   & float & Right ascension of the galaxy [degree]$^{b}$\\
  dec   & float & Declination of the galaxy [degree]$^{b}$\\
  redshift   & float   & Redshift$^{b}$\\
  effr\_frac   & float   & Galaxy's distance to the center of its host void, normalised to the effective radius of the void$^{a}$\\
  morphology   & float   & Morphology type$^{c}$\\
  d25   & float   & Major axis [arcsec]$^{a}$\\
  incl   & float   & Inclination [degree]$^{a}$\\
  pa   & float   & Position angle [degree]$^{d}$\\
  rabsmag   & float   & Absolute r-band magnitude [mag]$^{b}$ \\
  sb   & float   & Surface brightness [mag.arcsec$^{-1}$]$^{a}$\\
  g\_r & float & Color g-r [mag]$^{b}$\\
  Ar & float & Galactic r band extinction along line-of-sight [mag]$^{a}$\\
  M\_star & float & Log of stellar mass in solar masses $^{e}$ \\
  fluxHI & float & HI line flux density [Jy.km.s$^{-1}$]$^{f}$\\
  vmaxg & float & Rotational velocity [km.s$^{-1}$]$^{d}$\\
  logSFR & float & Log of star formation rate [log(M$_{\odot}$.yr$^{-1}$)]$^{e}$\\
  SDSS\_ID & float & The SDSS ID$^{g}$\\
  \hline
 \end{tabular}\\
 \textit{a}:\,\cite{Perez:2024},
\textit{b}:\,\cite{Pan:2012},
\textit{c}:\,\cite{2018MNRAS.476.3661D},
\textit{d}:\,\cite{Makarov:2014}
\\
\textit{e}:\, \href{https://live-sdss4org-dr12.pantheonsite.io/spectro/galaxy_mpajhu/}{The MPA-JHU survey},
\textit{f}:\,\cite{Durbala:2020},
\textit{g}:\,\hyperlink{https://www.sdss3.org/dr8/}{Public Sloan Digital Sky Survey Catalogue-DR8}
\\
\textit{*}Available on the project webpage
\end{table*}

\subsection{SQL query page}\label{sec:access3}
Users can access all the DR1 catalogs and CAVITY PPak data cubes published through the 'query' page (see Fig.~\ref{Landing_page}) by performing SQL or Astronomical Data Query Language (ADQL, IVOA standard language) queries. Narrowing down the sample selection by filtering or cross matching with multiple catalogs on the database or uploaded data is possible. Cross correlations with Simbad and VizieR services are also the available options on this page. It should be noted that keywords and unquoted identifiers are case insensitive in the SQL. Hence, in general, lower-case identifiers are used throughout the database, and ‘galaxy’ is the main identifier of target galaxies in this survey. 
An example of SQL request on how to select all the late type galaxies (morphological T-type greater than 0) at a redshift lower than 0.030 and belonging to voids 482 or 487 is outlined below:

\begin{lstlisting}[language=SQL]
SELECT galaxy, file FROM dr1.PPAK_cubes
WHERE galaxy IN (
    SELECT galaxy FROM dr1.Master
    WHERE morphology > 0 
    AND redshift <= 0.030 
    AND void IN (482, 487)
)
\end{lstlisting}

\section{Summary\label{sec:summary}}

In this article, we detail the data reduction, primary characteristics, and data access of the first public data release (DR1) from the Calar Alto Void Integral-field Treasury SurveY (CAVITY). DR1 includes 100 data cubes obtained with the V500 set-up of the PMAS/PPak spectrograph, covering a wavelength range of 3745–7500 \AA\ and offering a spectral resolution of 6.0 \AA\ (FWHM). This initial release features 100 void galaxies, which represent one-third of the total anticipated sample. These galaxies span a range of masses, morphological types, and colors similar to those of the full sample, except for a lack of low-mass systems due to the observational limitations of the instrument in the faint regime. The randomly selected subset the constraint of the evolution of galaxies residing in cosmic voids across the color–magnitude diagram down to -18 mag in $r$-band. CAVITY DR1 delivers science-grade, quality-checked integral-field spectroscopic data to the research community\footnote{\href{https://cavity.caha.es}{cavity.caha.es}}.

 We have outlined the main quality parameters analyzed during the validation process, providing users with complete tables to aid in selecting objects for their science cases. The data were reduced using the first official version of the CAVITY pipeline (V1.2), which operates automatically. We plan to enhance this pipeline and consistently offer re-reduced complete datasets to the community in the next data release.

 The quality of each data cube underwent cross-checking using a set of defined figures of merit, enabling us to select and distribute data of the highest quality. Parameters such as wavelength calibration accuracy, spectral resolution, sky subtraction, and the stability of observations and reduction processes were among those cross-checked.
 
 The data distributed in DR1 already include more than 100k independent spectra. While this constitutes a sizable sample, it represents only one-third of the total number of spectra anticipated upon completion of the CAVITY survey. Currently, data has been acquired for over 200 galaxies, for which we will continue to conduct the quality control tests implemented for this data release. It is anticipated that the second and final CAVITY data release will occur after 2026 and by that  time we aim to have collected high-quality science-grade data for more than 300 galaxies in voids.

\section*{Data availability}
The `master' table, QC tables, and catalogs description are available at: \urls{\urltabs}
   
\begin{acknowledgements}
Based on observations collected at the Centro Astron\'omico Hispano en Andaluc\'ia (CAHA) at Calar Alto, operated jointly by Junta de Andaluc\'ia and Consejo Superior de Investigaciones Cient\'ificas (IAA-CSIC),  under the CAVITY legacy project. We acknowledge financial support by the research projects AYA2017-84897-P, PID2020-113689GB-I00, and PID2020-114414GB-I00, financed by MCIN/AEI/10.13039/501100011033, the project A-FQM-510-UGR20 financed from FEDER/Junta de Andaluc\'ia-Consejer\'ia de Transformaci\'on Econ\'omica, Industria, Conocimiento y Universidades/Proyecto and by the grants P20-00334 and FQM108, financed by the Junta de Andaluc\'ia (Spain).
      RGB, AC, and RGD acknowledge financial support from the Severo Ochoa grant CEX2021-001131-S funded by MCIN/AEI/ 10.13039/501100011033 and to grant PID2022-141755NB-I00.
      LSM acknowledges support from Juan de la Cierva fellowship (IJC2019- 041527-I).
      TRL acknowledges support from Juan de la Cierva fellowship (IJC2020-043742-I).
      SDP acknowledges financial support from Juan de la Cierva Formaci\'on fellowship (FJC2021-047523-I) financed by MCIN/AEI/10.13039/501100011033 and by the European Union `NextGenerationEU'/PRTR, Ministerio de Econom\'ia y Competitividad under grants PID2019-107408GB-C44 and PID2022-136598NB-C32, and is grateful to the Natural Sciences and Engineering Research Council of Canada, the Fonds de Recherche du Qu\'ebec, and the Canada Foundation for Innovation for funding.
      BB acknowledges financial support from the Grant AST22\_4.4, funded by Consejería de Universidad, Investigación e Innovación and Gobierno de España and Unión Europea – NextGenerationEU, and by the research project PID2020-113689GB-I00 financed by MCIN/AEI/10.13039/501100011033.
      MAF acknowledges support from the Emergia program (EMERGIA20\_38888) from Consejer\'ia de Universidad, Investigaci\'on e Innovaci\'on de la Junta de Andaluc\'ia.
      SBD acknowledges financial support from the grant AST22.4.4, funded by Consejer\'ia de Universidad, Investigaci\'on e Innovaci\'on and Gobierno de Espa\~na and Uni\'on Europea –- NextGenerationEU, also funded by PID2020-113689GB-I00, financed by MCIN/AEI.
      DE acknowledges support from a Beatriz Galindo senior fellowship (BG20/00224) from the Spanish Ministry of Science and Innovation.
      GTR acknowledges financial support from the research project PRE2021-098736, funded by MCIN/AEI/10.13039/501100011033 and FSE+.
      MIR acknowledges financial support by the research projects PID2020-113689GB-I00, and PID2020-114414GB-I00, financed by MCIN/AEI/10.13039/501100011033, the project A-FQM-510-UGR20 financed from FEDER/Junta de Andaluc\'{\i}a-Consejer\'{\i}a de Transformaci\'on Econ\'omica, Industria, Conocimiento y Universidades/Proyecto and by the grants P20-00334 and FQM108, financed by the Junta de Andaluc\'{\i}a (Spain) and Grant AST22.4.4, funded by Consejer\'{\i}a de Universidad, Investigaci\'on e Innovaci\'on and Gobierno de Espa\~na and Uni\'on Europea – NextGenerationEU.
      HMC acknowledges support from the Institut Universitaire de France and from Centre National d’Etudes Spatiales (CNES), France. 
      JFB acknowledges support from the PID2022-140869NB-I00 grant from the Spanish Ministry of Science and Innovation.
      LG acknowledges financial support from AGAUR, CSIC, MCIN and AEI 10.13039/501100011033 under projects PID2020-115253GA-I00, PIE 20215AT016, CEX2020-001058-M, and 2021-SGR-01270.
      AFM acknowledges support from RYC2021-031099-I and PID2021-123313NA-I00 of MICIN/AEI/10.13039/501100011033/FEDER,UE,NextGenerationEU/PRT.
      IMC acknowledges support from ANID programme FONDECYT Postdoctorado 3230653 and ANID, BASAL, FB210003.
      Funding for this work/research was provided by the European Union (MSCA EDUCADO, GA 101119830). % Reynier
      PMSA acknowledges grants PID2019-105602GBI00 and PID2022-136505NB-I00.
      JR acknowledges ﬁnancial support from the Spanish Ministry of Science and Innovation through the project PID2022-138896NB-C55.
      PSB acknowledges support from grant  PID2022-138855NB-C31 funded by MCIN/AEI/ 10.13039/501100011033.
      PVG acknowledges that the project that gave rise to these results received the support of a fellowship from “la Caixa” Foundation (ID 100010434). The fellowship code is B005800.
      This research made use of astropy, a community-developed core python \citep[http://www.python.org,][]{python} package for Astronomy \citep{2013A&A...558A..33A,2018AJ....156..123A,2022ApJ...935..167A}; IPython \citep{PER-GRA:2007}; Matplotlib \citep{Hunter:2007}; NumPy \citep{2011CSE....13b..22V}; SciPy \citep{2020SciPy-NMeth,scipy_11255513}; Pandas \citep{mckinneyprocscipy2010} and TOPCAT \citep{2005ASPC..347...29T}.
\end{acknowledgements}

% Bibliography
\bibliographystyle{aa}
\bibliography{bib_cavity_dr1}

\end{document}